\documentclass{article}
\usepackage{arxiv}

\pdfoutput=1

\usepackage[utf8]{inputenc} 
\usepackage[T1]{fontenc}    
\usepackage{url}   
\usepackage{booktabs} 
\usepackage{amsfonts} 
\usepackage{nicefrac} 
\usepackage{microtype}
\usepackage{lipsum}
\usepackage{graphicx}
\graphicspath{ {./figures/} }
\usepackage[version=4]{mhchem} 
\usepackage{tabularx}

\usepackage{siunitx}
\usepackage{multirow}
\usepackage[table]{xcolor}
\usepackage{listings} 
\usepackage{xcolor} 
\usepackage{amsmath}
\usepackage{amssymb}
\usepackage{amsthm}
\usepackage{mathtools}
\usepackage{tikz}
\usetikzlibrary{matrix,chains,positioning,decorations.pathreplacing,arrows}
\usepackage{algorithm}
\usepackage[normalem]{ulem}
\usepackage[noend]{algpseudocode}
\usepackage{filecontents}
\usepackage{xr-hyper}
\usepackage{hyperref} 
\usepackage{cleveref}
\usepackage[numbers]{natbib}

\makeatletter
\newcommand*{\addFileDependency}[1]{
  \typeout{(#1)}
  \@addtofilelist{#1}
  \IfFileExists{#1}{}{\ypeout{No file #1.}}
}
\makeatother

\makeatletter
\newcommand*{\addAuxFileDependency}[1]{
  \makeatletter\@input{x#1.tex}\makeatother
}
\makeatother

\small

\numberwithin{equation}{section}
\numberwithin{equation}{subsection}

\algrenewcommand{\algorithmicrequire}{\textbf{Input:}}
\algrenewcommand{\algorithmicensure}{\textbf{Output:}}

\crefname{equation}{Eq.}{Eqs.}
\Crefname{equation}{Equation}{Equations}
\crefname{table}{Table}{Tables}
\Crefname{table}{Table}{Tables}
\crefname{figure}{Fig.}{Figs.}
\Crefname{figure}{Figure}{Figures}

\title{Fitting micro-kinetic models to transient kinetics of temporal analysis of product reactors using kinetics-informed neural networks}

\author{
  Dingqi Nai\textsuperscript{a}, Gabriel S. Gusmão\textsuperscript{a}, Zachary A. Kilwein\textsuperscript{a}, Fani Boukouvala\textsuperscript{a}, Andrew J. Medford\textsuperscript{a}$^*$\\\\
  \textsuperscript{a}School of Chemical \& Biomolecular Engineering\\
  Georgia Institute of Technology\\
  Atlanta, GA 30332\\
}

\begin{document}

\maketitle

\let\thefootnote\relax\footnotetext{$*$~Corresponding author. Email: ajm@gatech.edu}

\begin{abstract}

The temporal analysis of products (TAP) technique produces extensive transient kinetic data sets, but it is challenging to translate the large quantity of raw data into physically interpretable kinetic models, largely due to the computational scaling of existing numerical methods for fitting TAP data. In this work, we utilize kinetics-informed neural networks (KINNs), which are artificial feedforward neural networks designed to solve ordinary differential equations constrained by micro-kinetic models, to model the TAP data. We demonstrate that, under the assumption that all concentrations are known in the thin catalyst zone, KINNs can simultaneously fit the transient data, retrieve the kinetic model parameters, and interpolate unseen pulse behavior for multi-pulse experiments. We further demonstrate that, by modifying the loss function, KINNs maintain these capabilities even when precise thin-zone information is unavailable, as would be the case with real experimental TAP data. We also compare the approach to existing optimization techniques, which reveals improved noise tolerance and performance in extracting kinetic parameters. The KINNs approach offers an efficient alternative for TAP analysis and can assist in interpreting transient kinetics in complex systems over long timescales.

\keywords{Physics-informed neural networks; Scientific machine learning; Kinetic modeling; Catalysis}

\end{abstract}

\section{Introduction} \label{s:intro}

Catalysis is a key component of the current chemical industry, and is involved in the production processes of more than 80\% of the items used in our daily life \citep{Morgan2017FortyProducts, Rasmussen2012MonitoringCatalyst, Thomas1997PrinciplesCatalysis}. Furthermore, catalysis will play an important role in the development of new chemical processes that improve energy production efficiency, flexible conversion of renewable feedstocks, and control of pollution emissions \citep{Anastas2002OriginsChemistry, Sheldon2007GreenCatalysis, Cavani2009SustainabilityEvolution, Gong2014CatalysisEnergy, Thomas2014HeterogeneousFeedstocks, Guo2014RecentChemistry, Papanikolaou2022CatalysisElectrocatalysis}. Therefore, the discovery and optimization of catalytic materials is an essential factor in the development of technologies to improve the economic and environmental sustainability of modern society \citep{Felpin2008HeterogeneousSustainability, Mitchell2020NanoscaleTechnologies}. As multicomponent solids, the effectiveness of many heterogeneous catalysts depends on their unique properties and operating conditions, such as particle size, chemical composition, and kinetic characteristics \citep{Chorkendorff2003ConceptsKinetics, Galhenage2014UnderstandingReaction, Gleaves2010TemporalCatalysts, Jinnouchi2017PredictingAlgorithm}. A slight change in the intrinsic performance of the catalyst can lead to a considerable change in conversion, yield, and selectivity \citep{Hagen2015IndustrialApproach, Morgan2016EvolutionReactions, Rothenberg2008Catalysis:Applications}. Therefore, a complete understanding of the properties of the catalyst and their relationship to the intrinsic kinetics of the elementary steps is desired to help rationally develop and optimize new catalyst materials \citep{Gleaves1997TAP-2:Approach, Matera2019ProgressCatalysis, Perez-Ramirez2007EvolutionTechnique}. 

Over the years, many techniques for obtaining the catalyst properties and intrinsic kinetics of catalyst materials have been continuously developed \citep{Eyring2004TheReactions, Evans1935SomeSolution, Kobayashi1972ApplicationTemperatures, Kobayashi1974TRANSIENTCATALYSIS, Bennett1976TheCatalysis, Biloen1983TransientMethods, Laidler1983TheTheory, Dumesic1993TheCatalysis, Kohn1996DensityStructure, DeBellefon1998AChromatograph, Cortright2001KineticsSchemes, Bligaard2004TheCatalysis, ANDERSSON2006TowardCatalysts, Nrskov2009TowardsCatalysts, Wang2011UniversalMetals, Wang2011UniversalReactions, Rangarajan2012Language-orientedRING, Hoffmann2014Kmos:Framework, Medford2015CatMAP:Trends, Greeley2016TheoreticalDesign, Goldsmith2017AutomaticCatalysis}. Catalytic experiments can be broadly divided into steady-state studies and transient kinetic studies. Steady-state techniques operate under the assumption that systems are designed to reach a steady state where the rate of change in reactor properties is effectively zero \citep{Jandeleit1998CombinatorialCatalysis, Shekhtman1999Thin-zoneApplication}. Steady-state operation allows simplification of the governing equations, but the experiments are time-consuming, and the low temporal resolution makes steady-state studies best for providing a perspective on the behavior of the catalyst under a single set of operating conditions. Moreover, the kinetic behavior of steady-state data is dominated by the rate-limiting steps, making it difficult to provide sufficient knowledge of each elementary step to predict how the catalyst will behave under a wide range of operating conditions \citep{Gleaves1997TAP-2:Approach, Perez-Ramirez2007EvolutionTechnique, Caravieilhes2002TransientMeasurements}.

In contrast to steady-state studies, transient kinetic studies function under designated disturbances, such as rapid changes in gas flow rate or temperature \citep{Morgan2017FortyProducts, Olea2009TemporalReactions, Yablonsky2003TemporalTheory, Gleaves1988TemporalResolution}. The dynamic response of the system to the induced time-dependent disturbance is then monitored using a mass spectrometer\citep{Yablonsky2016Rate-ReactivityCatalysts}. Even if short-lived intermediates are included, the high temporal resolution allows a more comprehensive understanding of each elementary step that controls the process, which is generally unavailable through steady-state studies \citep{Gleaves1997TAP-2:Approach, Yablonsky2003TemporalTheory, Biloen1983TransientMethods}. In particular, temporal analysis of products (TAP) is a well-studied transient kinetics tool \citep{Gleaves1988TemporalResolution, Gleaves1997TAP-2:Approach, Shekhtman1999Thin-zoneApplication, Constales2001Multi-zoneEquation}, which has attracted more attention in recent years due to the high volume of data and the ability to directly study powdered or supported catalysts \citep{Yablonsky2016Rate-ReactivityCatalysts, Yonge2021TAPsolver:Experiments, Yonge2022QuantifyingParameters, Redekop2014ElucidatingData, Redekop2022AligningCatalysts, Kunz2020ProbabilityExperiments, Kunz2021DataLearning, Batchu2020EthanolProducts}. TAP experiments generate rich data sets that provide fundamental quantitative insights into the material properties and transient kinetic responses of complex supported catalysts. However, converting the collected data into physically interpretable kinetic information is still challenging, particularly for multi-pulse datasets that may include hundreds or thousands of individual TAP pulses that gradually alter the state of the catalyst \citep{Gleaves1997TAP-2:Approach}.

Several steps have been taken to learn kinetic models and parameters from TAP data, including the derivation of the governing partial differential equations (PDEs) and PDE-constrained optimization, but some gaps still need to be filled \citep{Rothaeme1996ModelingReactor, VanDerLinde1997MathematicalEquations, Yablonskii1998Moment-basedExperiment, Delgado2002ModelingReactor, Kondratenko2008MechanisticDecomposition, Balcaen2009TransientO2, Balcaen2011KineticCatalysts, Kumar2011MicrokineticProducts, Roelant2011MathematicalExperiments, Redekop2013MomentarySites, Golman2016TransientPackage, Yablonsky2016Rate-ReactivityCatalysts, Reece2017KineticCatalysis, Batchu2020EthanolProducts, Kunz2021DataLearning, Yonge2021TAPsolver:Experiments}. Most of these tools exhibit a trade-off between scalability and model complexity and struggle to handle large reaction networks or multi-pulse TAP data efficiently. In other words, the extraction of kinetic information is still limited by the ``data velocity'' and the ``data volume'' of the ``V's'' of data science \citep{Medford2018ExtractingInformatics, Saggi2018AValue-creation, Chang2019NISTDefinitions}. To combat this issue, we introduce kinetics-informed neural networks (KINNs), a machine learning-based method, into the field of TAP data analysis. KINNs can fit the existing data while solving and parameterizing a physically interpretable kinetic model \citep{Gusmao2022Kinetics-informedNetworks}. The machine learning structure provides high scalability, enabling KINNs to handle multi-pulse data efficiently through a single network, even for reaction networks with many elementary steps. Additionally, the flexible loss function allows for the incorporation of information from different sources, which fully exploits the vast data sets that can be obtained through TAP experiments.  

This work demonstrates the theoretical and practical validity of using KINNs to extract kinetic information from TAP data sets. We present three case studies: ideal single-pulse, ideal multi-pulse, and practical multi-pulse. In all cases, simulated synthetic data is used so that the ground truth is known, although the rate constants and noise levels are inspired by an experimental data set to ensure that they are representative \citep{Yonge2021TAPsolver:Experiments}. In ideal cases, explicit reactor information from the catalyst zone (concentration and net flux) is available from the simulation. In contrast, the practical case mirrors a realistic scenario, where we assume that only fluxes at the reactor outlet are known and utilize data preprocessing in conjunction with the KINN formalism to estimate concentrations in the catalyst zone. The performance of KINNs is compared to the optimization via collocation in the case of the practical multi-pulse data. The results indicate that KINNs are a promising option for analyzing transient kinetics, particularly in the case of large or noisy datasets, and indicate that KINNs provide a scalable and flexible foundation for the development of new techniques in the analysis of TAP and other types of transient kinetic data.

\section{Methodology} \label{s:methods}

\subsection{Kinetics-informed neural networks}
\label{subsec:methods.1}

KINNs, developed by \citet{Gusmao2022Kinetics-informedNetworks}, solve ordinary differential equations (ODEs) utilizing feedforward neural nets (NNs) as a basis function \citep{Meade1994TheNetworks}. While training to fit concentration data, the NNs are also constrained by the micro-kinetic models (MKMs) that are in the form of ODEs \citep{Gusmao2015ASystems, Dumesic1993TheCatalysis}. At each time $t$, the rate of change of concentration based on the MKM is given as:
\begin{equation} \label{eqn:MKMs}
    \dot{\mathbf{c}} =\mathbf{M}\times\mathbf{k}(\theta) \circ \psi(\mathbf{c}) + \mathbf{f},
\end{equation}
where $\mathbf{M}\in\mathbb{R}^{n\times m}$ is the stoichiometriy matrix, and $\psi$ is a function that maps concentration $\mathbf{c}\in\mathbb{R}^n$ to the corresponding concentration-based power-law kinetic model. $\mathbf{k}\in\mathbb{R}^m$ is the temperature ($\theta$) dependent Arrhenius-like rate constant term, and $\mathbf{f}\in\mathbb{R}^n$ is the transport term. Here, $n$ represents the number of species, $m$ is the number of elementary reactions, and $\circ$ represents an element-wise product. A more detailed description of the formulation of KINNs can be found in previous work by \citet{Gusmao2022Kinetics-informedNetworks}. In this study, under the isothermal assumption, we simplify the rate law term in Eq. \ref{eqn:MKMs} to
\begin{equation} \label{eqn:MKMs_simp}
     \mathbf{r}(\mathbf{c}, \mathbf{k}) = \mathbf{M}\times\mathbf{k} \circ \psi(\mathbf{c})
\end{equation}
where $\mathbf{r}\in\mathbb{R}^n$.

During training, the residuals of the states and derivatives $\varepsilon$ are calculated as shown in Eq. \ref{eqn:residual}.
\begin{equation} \label{eqn:residual}
    \begin{gathered}
        \varepsilon_{c_i} = c_i(t, \boldsymbol{\omega_s}) - \tilde{c_i} \\
        \varepsilon_{\dot{c}_i} = \frac{dc_i(t, \boldsymbol{\omega_s})}{dt} - (r_i(\mathbf{c}(t, \boldsymbol{\omega_s}),\mathbf{k}) + f_i)
    \end{gathered}
\end{equation}
Here the tilde term represents the target value, and therefore the loss function $J$ is defined as the combination of data loss $j_{data}$ and MKM loss $j_{model}$ using a weighted parameter $\alpha$, which follows Eq. \ref{eqn:loss_fn} \citep{Gusmao2022Kinetics-informedNetworks}.
\begin{equation} \label{eqn:loss_fn}
    \begin{gathered}
    \min_{\boldsymbol{\omega}} \quad J(\boldsymbol{\omega}) = j_{data}(\boldsymbol{\omega_s}) + \alpha j_{model}(\boldsymbol{\omega_s},\mathbf{k}) \\
    j_{data}(\boldsymbol{\omega_s}) = \sum_i^n\varepsilon_{c_i}^T\varepsilon_{c_i} \\
    j_{model}(\boldsymbol{\omega_s},\mathbf{k}) = \sum_i^n\varepsilon_{\dot{c}_i}^T\varepsilon_{\dot{c}_i}
    \end{gathered}
\end{equation}
The tuple $\boldsymbol{\omega} = (\boldsymbol{\omega_s},\mathbf{k})$ combines the NN parameters, or the weights of the connections, $\boldsymbol{\omega_s}$ and model (kinetic) parameters $\mathbf{k}$. In the forward process, the Adam optimizer minimizes data loss $j_{data}$ in the least squares sense. During training, the Adam optimizer \citep{Kingma2014Adam:Optimization} is also used to solve the inverse problem with the same network, minimizing the MKM loss, $j_{model}$ \citep{Raissi2019Physics-informedEquations}.

This loss function grants KINNs significant scalability and flexibility in the sources of information it can handle, since the neural network acts as an interpolator that can smooth noise or fill in missing values. The approach scales  well to large reaction networks, since the kinetic parameters enter the optimization in the same way as NN weights, so that the size of the reaction network becomes limiting only if the number of elementary steps approaches the number of parameters in the NN (typically $>100$). In addition, other information, such as reaction thermodynamics, DFT energies, or transient spectroscopic signals, can, in principle, be included as additional constraints to MKMs by including more residual components in the loss $J$ with proper weighting hyperparameters. The relative focus of the network can be altered between the model and the data by adjusting the hyperparameter $\alpha$. Unfortunately, results can be sensitive to this choice, and there is currently no standardized method to choose an optimum $\alpha$; instead, this must be done manually using previous knowledge of the relative importance between the data and the model, cross-validation techniques, or multi-objective optimization to identify an optimal trade-off \citep{Gusmao2022Kinetics-informedNetworks}. In the field of power system engineering, some works have employed Lagrangian duality to iteratively adjust the hyperparameter values \citep{Fioretto2019PredictingMethods, Jalving2024Physics-informedFlow}. Although these works are not targeted to kinetic modeling, they may still offer useful insights for hyperparameter tuning. Recently, the development of robust KINNs has resolved this issue by using a maximum-likelihood estimation formalism \citep{Gusmao2023Maximum-likelihoodProblems}. However, the method assumes that all intermediate states are known, so it is not directly applicable to TAP data, although future work will focus on adaptation of robust KINNs to the TAP problem.

\subsection{Transport and kinetic model in TAP reactor}
\label{subsec:methods.2}

A typical TAP system consists of a catalyst zone sandwiched between two inert zones, and the reactor operates under isothermal vacuum conditions. During experiments, a series of nanomolar pulses of reactant/inert gases are injected into the reactor, and a mass spectrometer is placed at the outlet to instantly quantify the outlet flow of gas molecules with a millisecond time resolution. Gas transport occurs in the Knudsen diffusion regime, which provides well-defined transport properties for the reactor \citep{Morgan2017FortyProducts, Gleaves1988TemporalResolution, Gleaves1997TAP-2:Approach, Yablonsky2003TemporalTheory}. A more detailed explanation of the TAP reactor is beyond the scope of this paper and can be found in previous work including \citet{Gleaves1988TemporalResolution} and others.

The one-dimensional model that describes the dynamics, including both diffusion and chemical reaction, inside the TAP reactor is
\begin{equation} \label{eqn:govern_eqn}
    e \frac{\partial c_{i}}{\partial t} =  \frac{\partial}{\partial x} \left( \mathcal{D}_i \frac{\partial c_{i}}{\partial x}\right) + r_i(\mathbf{c}, \mathbf{k}),
\end{equation}
where $c_{i}$ stands for the concentration for species $i$ at location $x$ at time $t$, $\mathcal{D}$ is effective Knudsen diffusion coefficient, $e$ is bed voidage, and $r_i$ is the reaction rate for species $i$, which is a function of gas phase species and adspecies concentrations in catalyst zone and intrinsic kinetic parameters \citep{Yablonsky2007TheModel}. In practice, the catalyst zone is often purposefully designed to be much thinner than the inert zones, following the thin-zone TAP reactor (TZTR) proposed by \citet{Shekhtman1999Thin-zoneApplication}. The TZTR ensures a highly uniform chemical composition and stable catalyst properties during experiments, so the concentration gradient inside the catalyst zone can be neglected. Under these assumptions, the thin catalyst zone can then be described as a diffusional transient CSTR \citep{Shekhtman1999Thin-zoneApplication, Constales2001Multi-zoneEquation, Constales2017PreciseAssumptions}. In this case, the diffusion term in the above governing equation can be approximated to
\begin{equation} \label{eqn:diffusion_approx}
    \frac{\partial}{\partial x} \left( \mathcal{D}_i \frac{\partial c_{i}}{\partial x}\right) \approx \frac{f_{i}^{in}(t) - f_{i}^{out}(t)}{\mathit{l_{cat}}},
\end{equation}
where $f_{i}^{in}$ and $f_{i}^{out}$ are the flux into and out of the thin zone for species $i$, and $\mathit{l_{cat}}$ is the catalyst zone length. So, when the explicit flux at each timestamp is available, the spatial dependence in Eq. \ref{eqn:govern_eqn} can be eliminated, and the PDE governing equation, Eq. \ref{eqn:govern_eqn}, can be approximated to an ODE as
\begin{equation} \label{eqn:ode_govern}
    e\frac{dc_{i}}{dt} = \frac{f_{i}^{in} - f_{i}^{out}}{\mathit{l_{cat}}} + r_i(\mathbf{c}, \mathbf{k}),
\end{equation}
which can be directly mapped to the mathematical form of KINNs in Eq. \ref{eqn:MKMs_simp}.

\subsection{Data generation and processing} 
\label{subsec:methods.3}

Carbon monoxide oxidation is a common benchmark reaction in the field of catalysis and TAP experiments \citep{Redekop2014ElucidatingData}, and we adopt it as a case study here to support the validity of the KINNs approach. The elementary steps and kinetic parameters are provided in Table \ref{tab:elementary_rxns}, which were generated from previous TAPSolver simulations and serve as the ground truth in this study \citep{Yonge2021TAPsolver:Experiments}. Additional information about TAPSolver can be found in previous work by \citet{Yonge2021TAPsolver:Experiments} and scripts to reproduce the data are provided in the Supplementary Information. TAPSolver simulations provide concentrations in the catalyst zone as well as the net flux, which can then be directly fed into KINNs --- this is referred to as the ``ideal case'' since the exact concentrations of all species (including adspecies) in the thin zone are assumed to be known. In the ideal case, we omit noise from the data to evaluate the ability of KINNs to recover kinetic parameters. The robustness of KINNs to noise is explicitly demonstrated using the practical multi-pulse data set in Section \ref{subsec:results.3} and the performance comparison in Section \ref{subsec:comp_pyomo}. 

\begin{table}[h]
    \centering
    \begin{tabularx}{\textwidth}{c @{\extracolsep{\fill}} c c c}
    \toprule
        Kinetic parameter & Reaction & Actual value & Units \\
    \midrule
        $k_{1}$  &  CO + $*$          $\;\to\;$  CO$*$           & $15.0$     & $\frac{cm^3}{nmols}$   \\[0.3cm]
        $k_{-1}$ &  CO$*$             $\;\to\;$  CO + $*$        & $0.70$     & $\frac{1}{s}$          \\[0.3cm]
        $k_{2}$  &  O$_2$ + 2$*$      $\;\to\;$  2O$*$           & $0.33$     & $\frac{cm^6}{nmol^2s}$ \\[0.3cm]
        $k_{-2}$ &  2O$*$             $\;\to\;$  O$_2$ + 2$*$    & -          & $\frac{cm^3}{nmols}$   \\[0.3cm]
        $k_{3}$  &  CO$*$ + O$*$      $\;\to\;$  CO$_2$ + 2$*$   & $0.40$     & $\frac{cm^3}{nmols}$   \\[0.3cm]
        $k_{-3}$ &  CO$_2$ + 2$*$     $\;\to\;$  CO$*$ + O$*$    & $0.02$     & $\frac{cm^6}{nmol^2s}$ \\[0.3cm]
        $k_{4}$  &  CO + O$*$         $\;\to\;$  CO$_2$ + $*$    & $15.2$     & $\frac{cm^3}{nmols}$   \\[0.3cm]
        $k_{-4}$ &  CO$_2$ + $*$      $\;\to\;$  CO + O$*$       & -          & $\frac{cm^3}{nmols}$   \\[0.3cm]
    \bottomrule
    \end{tabularx}
    \caption{List of elementary steps and kinetic parameters used to generate the synthetic carbon monoxide oxidation data. Dashed value indicates the parameter was not included in simulation.}
    \label{tab:elementary_rxns}
\end{table}

\begin{figure}[h]
    \centering\includegraphics[keepaspectratio=true,width=0.82\textwidth]{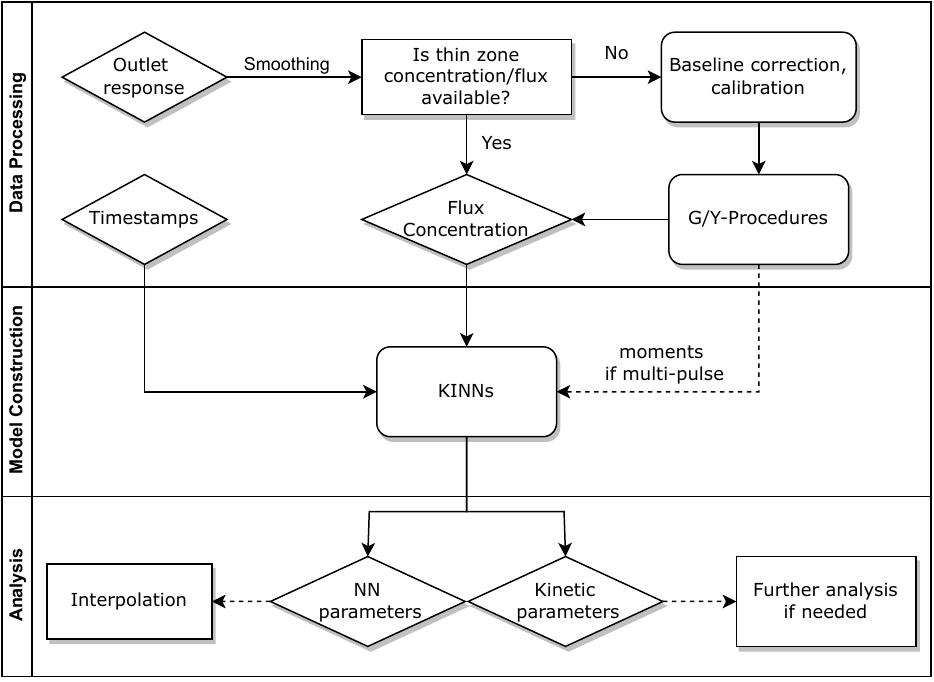}
    \caption{Workflow of modeling TAP data with KINNs. The raw outlet response is processed to obtain thin zone concentration and flux that are fed into constructed KINN model. The obtained KINN parameters can be used to interpolate single or multi-pulse data, and the kinetic parameters can be used directly in kinetic models or serve as an initial guesses for refinement using PDE tools like TAPSolver.}
    \label{fig:workflow}
\end{figure}

In contrast to simulated data, in real experiments, the catalyst zone concentration and fluxes are not directly measurable, and noise will be present. Therefore, noise was added as a Gaussian error distribution centered around 0 with the width of half the standard deviation of estimated signals. TAP noise was shown to be a combination of Gaussian noise and spectrally localized noise, and Gaussian noise was found to be the dominant noise source in many data sets \citep{Roelant2007NoiseResponses}.The noise level is proportional to the intensity of the signal and the noise is heteroskedastic, which means that the proportionality factor varies over different data. However, here we assume a constant proportionality factor as a first approximation. To estimate the catalyst zone information in the realistic scenario, the Y-procedure is applied to approximate thin zone concentrations from the outlet flux data. The Y-procedure was introduced by \citet{Yablonsky2007TheModel}. This technique provides a route for extracting the reaction rate and concentration of gas species from the secondary TAP response outlet gas flux, without any prior assumptions about the kinetic or diffusion models \citep{Kunz2020ProbabilityExperiments, Yablonsky2007TheModel}. \citet{Yablonsky2007TheModel} showed that when the thin zone assumption holds, the difference of two fluxes is equal to the estimated reaction rate $\mathbf{r}^{Y}$, as
\begin{equation} \label{eqn:rxn_rate}
    \mathbf{f}^{in}(t) - \mathbf{f}^{out}(t) = \mathbf{r}^{Y}(t).
\end{equation}
The concentration of adspecies is not directly available through the Y-procedure, but the ``atomic uptakes'', which are defined as the amount of each elemental species present on the surface, can be added to the loss function in Eq. \ref{eqn:loss_fn} to constrain the concentration of adspecies in the thin zone. Gas uptake $U$ is determined as the integral of each reaction rate with respect to time scaled by stoichiometry as:
\begin{equation} \label{eqn:uptake}
    U_i = \int_0^t \sum_i(\nu_i r_i^+ - \nu_i r_i^-)dt.
\end{equation}
where $\nu$ represents the stoichiometric coefficients \citep{Kunz2021DataLearning,Yablonsky2007TheModel, Redekop2011TheAdsorption, RossKunz2018PulseApproach}. By using the results of noisy TAPSolver simulations to calculate estimated thin-zone gas concentration, net flux, and surface uptake, the synthetic data become a direct surrogate for experimental data that can be fed into KINNs to derive the intrinsic kinetics. Fig. \ref{fig:workflow} shows a typical workflow for this analysis.

Before feeding the data into the KINN model, the data was subsampled non-uniformly to have a higher density of data points in the high-derivative region. Specifically, half of the data points were taken from the 0.0 to 0.5 $s$ interval. Additional sampling techniques, such as uniform sampling in log space and Chebyshev sampling, could be considered, but exploring the optimal sampling technique is beyond the scope of this paper. Concentrations and uptake are normalized to a range between 0 and 1 to ensure numerical stability and efficiency within the NN. Moreover, instead of using raw timestamps $t$, we transform the timestamps with a natural logarithm to enhance the resolution of the KINN at short timescales, which often features rapid changes in concentration. This logarithmic transformation ensures that the KINN can adapt more effectively to these early, rapidly evolving dynamics.

\subsection{KINNs setup} 
\label{subsec:methods.4}

To model the single-pulse TAP response, we constructed a 2-layer 8 hidden unit KINN (70 adjustable parameters) with a swish activation function \citep{Ramachandran2017SearchingFunctions}, where the input is the time representation and the outputs are the scaled thin zone concentrations. During training, $\alpha$ was initially set to $1\times10^{-10}$, which implied that the network would fit the concentrations without taking the MKMs into account. After every 5 epochs, each of which had 1000 iterations, $\alpha$ was increased by a factor of 10 until it reached 1. Then, $\alpha$ could be increased or decreased, and the direction and amount are manually adjusted based on the relative value between $j_{data}$ and $j_{model}$, and the parity behavior between the predicted and target values. Switching the focus back and forth effectively alters the initial guess of the NN weights for each value of $\alpha$, and allows KINNs to find NN weights that optimally fit both the MKMs and the concentrations. For $\alpha$ tuning, we rely on the loss value and $r^2$ correlation for both the concentration data and the kinetic model. This approach is chosen because the model residual, as defined in Eq. \ref{eqn:residual}, is sensitive to changes in the predicted concentration values. Specifically, adjustments to $j_{data}$ can negatively impact the fit of the kinetic model by affecting $j_{model}$, even if they result in a lower overall loss value. Monitoring these factors independently provides a route to balance the trade-offs between data fidelity and model accuracy. A more systematic optimization of $\alpha$ is beyond the scope of this work, but more detail can be found in previous work by \citet{Gusmao2022Kinetics-informedNetworks}.

Due to the increased size and complexity of multi-pulse dataset, we increased the neural network's complexity. For multi-pulse fitting, we constructed a 3-layer KINN with 10 hidden units in each layer, resulting in a total of 226 adjustable parameters. The zeroth moments ($m_{0}$) of the primary TAP response outlet flux of a given pulse were included as an additional input. The $m_{0}$ of the outlet flows is equal to the amount of corresponding gas species that exit the reactor during the specific pulse \citep{Kunz2020ProbabilityExperiments, Kunz2022InternalLearning, Yablonskii1998Moment-basedExperiment}; therefore, using it as an additional input provides the KINN with information about the surface conditions during the specific pulse and indicates the relative position of the current pulse in a series of pulses. To test KINN's ability to interpolate unseen pulses, which is critical to sparsify dense TAP data consisting of hundreds or even thousands of pulses, we used a synthetic 10-pulse CO oxidation data set, where pulses 0, 1, 2, 5, 8 were used for training, and pulses 3, 4, 6, 7, 9 served as testing sets.

The kinetic parameters are treated as part of KINNs' parameter set,  allowing their uncertainty to be estimated through the inverse of the Hessian matrix of the loss function as
\begin{equation} \label{eqn:inv_hessian}
    \mathbf{P} = \mathbf{H}_{J}^{-1}(\mathbf{k}),
\end{equation}
where $\mathbf{P}$ represents the inverse of Hessian of the loss function, $\mathbf{H}_{J}$, with respect to the kinetic parameters $\mathbf{k}$. The standard deviation $\sigma_i$ of the parameter $i$ can be calculated by taking the square root of the diagonal values using Eq. \ref{eqn:std} 
\begin{equation} \label{eqn:std}
    \sigma_{i} = \sqrt{P_{i,i}}
\end{equation}
based on the parameter's covariance \citep{Yonge2021TAPsolver:Experiments, Franceschini2008Model-basedArt, Zhan2021UncertaintyModels}. This is not a rigorous error estimate because the Hessian, and therefore the confidence intervals, depend on the  hyperparameters that control the relative weights in the loss function \citep{Gusmao2023Maximum-likelihoodProblems}. However, the estimate is still useful since it provides a semi-quantitative sensitivity analysis to each kinetic parameter.

\section{Results and Discussion} 
\label{sec:results}

To assess the ability of KINNs to extract kinetic parameters from transient kinetics measured with a TAP reactor, we conducted three case studies, including single-pulse CO oxidation with explicit thin zone information (``single-pulse ideal scenario''), 10-pulse CO oxidation with explicit thin zone information (``multi-pulse ideal scenario''), and 10-pulse CO oxidation with noise and fluxes and uptakes estimated using the Y-procedure (``multi-pulse practical scenario''). Finally, we compare the results of the multi-pulse noisy data fitting to the results obtained using collocation, a standard non-linear optimization technique for finding parameters of ODEs from data.

\subsection{Single-pulse ideal scenario}
\label{subsec:results.1}

\begin{figure*}[h]
    \centering\includegraphics[keepaspectratio=true,scale=0.6]{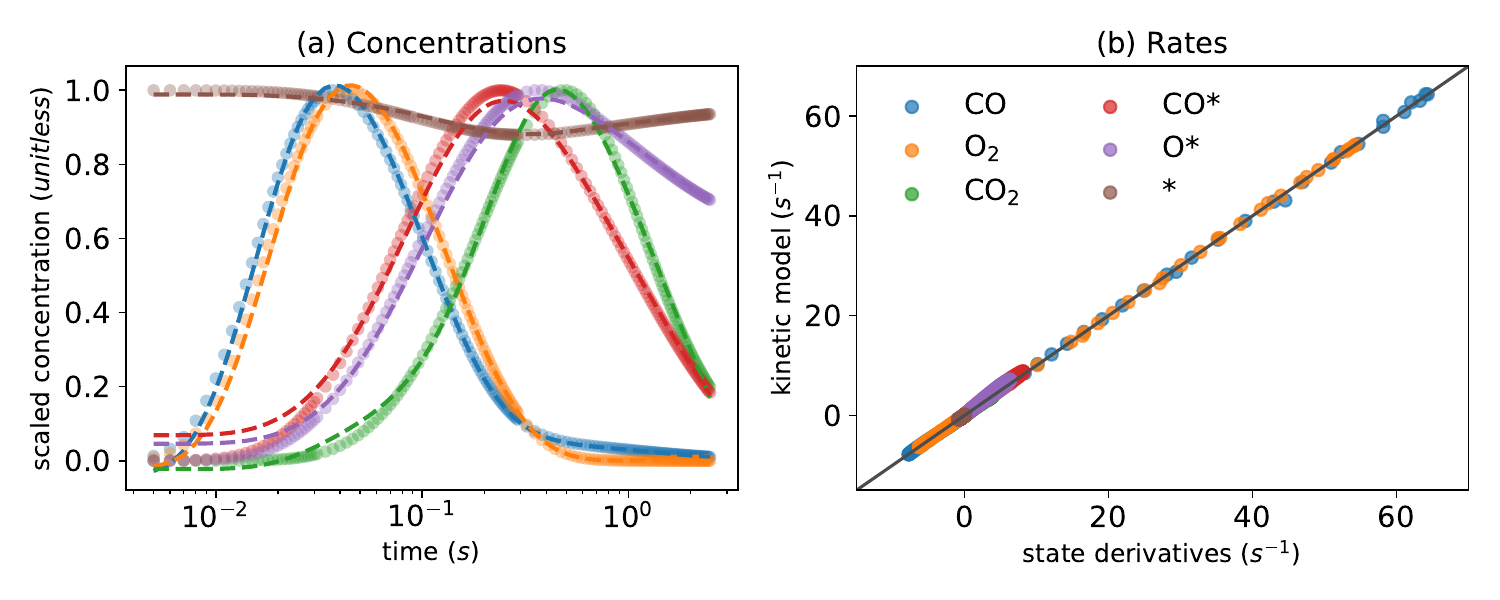}
    \caption{KINN optimization of single-pulse CO oxidation under minimization of cost function $J$ in Eq. \ref{eqn:loss_fn} for (a) concentrations, where circles represent target values and dashed lines represent predicted values, and (b) a parity plot of rates extracted from the KINN model and predicted from the kinetic model.}
    \label{fig:CO_single}
\end{figure*}

Fig. \ref{fig:CO_single}(a) and (b) show the performance of the KINN for predicting the concentrations and rates, respectively. The predicted concentrations from the KINN matches the target data for the scaled concentration well, achieving a Mean Absolute Error (MAE) of $1.39\times10^{-2}$, which is equivalent to $4.66\times10^{-2}$ $nmol/cm^3$ in the original scale. A minor discrepancy in the initial concentration of the adspecies could be further refined by incorporating constraints on the initial condition into the loss function \citep{Gusmao2022Kinetics-informedNetworks}. However, we omit this step since the goal of this study is to work toward realistic multi-pulse scenarios where initial conditions may not be known exactly. Furthermore, the kinetics based on the estimated concentrations demonstrate a clear parity with the automatic differentiated derivatives, as indicated by an MAE of $4.20\times10^{-1}$ $nmol/cm^3/s$ and $r^2$ values of $\sim$ 0.99 for both concentrations and rates.

\begin{figure}[h]
    \centering\includegraphics[keepaspectratio=true,scale=0.5]{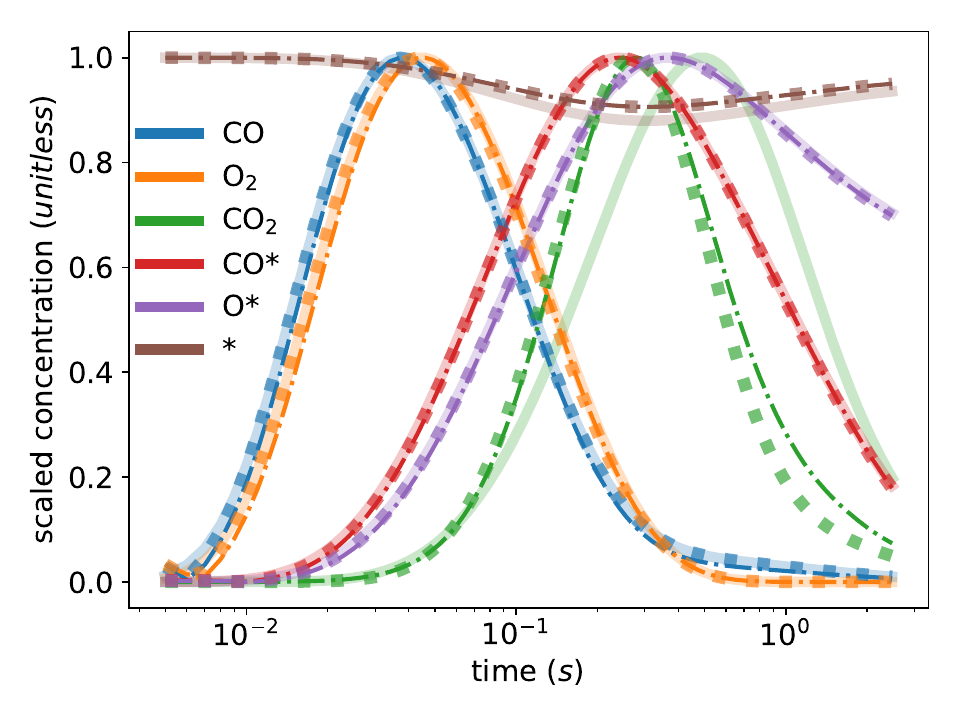}
    \caption{Ground truth TAP curve (solid line) and TAP curve obtained from solving ODEs with kinetic parameters extracted from the KINN (dash-dotted line) and the ground truth parameters (dotted line).}
    \label{fig:single_rebuild}
\end{figure}

Table \ref{tab:pred_params} presents the values of the converged kinetic parameters in this case study. The initial guess for all parameters was $1\times 10^{-5}$, which is far from the true values to demonstrate the robustness of the KINN and to avoid bias during optimization. The agreement between the converged and true values illustrates the ability of KINNs to extract intrinsic kinetic parameters when explicit concentration and flux are utilized. Fig. \ref{fig:single_rebuild} compares the ground truth TAP curve and the TAP curve obtained by solving the ODE model, using both the KINN-extracted and ground truth kinetic parameters. For all species except $CO_2$, the solved concentration profiles are nearly indistinguishable. The discrepancy observed in $CO_2$ is mainly due to the approximation of a PDE-based to an ODE-based system, as evidenced by the fact that the two curves derived from ODEs are very similar. A possible reason we did not observe this discrepancy for the reactants is that the reactant gases are present from the beginning of the experiment, and therefore their consumption rates are more straightforward to model accurately. In contrast, $CO_2$, the product, is formed as the experiment progresses, and its formation rate can be influenced by more factors. This complexity likely enlarges the discrepancy in its approximation. Additionally, the slower diffusion rate of $CO_2$ might render the transport term dominant more in its gradient. Consequently, omitting the rigorous transport model may contribute significantly to the deviation in its predicted concentration profile. Despite these discrepancies, the predicted kinetic parameter values are within an order of magnitude of the ground truth parameters, and the MAE for the apparent activation barriers, calculated using the Eyring equation, is $0.019$ eV, which is an order of magnitude lower than typical DFT errors. These results indicate that the ODE-based KINNs approach is capable of accurately recovering kinetic parameters from TAP data.

\begin{table*}[h]
\centering
\begin{tabularx}{\textwidth}{c @{\extracolsep{\fill}} c c c c}
    \toprule
        Parameter & Actual value & Single-pulse ideal & Multi-pulse ideal & Multi-pulse practical \\
    \midrule
        $k_{1}$   &  $15.0$      &  $11.5 \pm 0.091$  & $11.5 \pm 0.120$  &  $6.36 \pm 0.242$ \\
        $k_{-1}$  &  $0.70$      &  $0.53 \pm 0.048$  & $0.52 \pm 0.061$  &  $0.69 \pm 0.334$ \\
        $k_{2}$   &  $0.33$      &  $0.25 \pm 0.003$  & $0.25 \pm 0.004$  &  $0.14 \pm 0.008$ \\
        $k_{3}$   &  $0.40$      &  $0.37 \pm 0.029$  & $0.31 \pm 0.041$  &  $0.44 \pm 0.183$ \\
        $k_{-3}$  &  $0.02$      &  $0.04 \pm 0.010$  & $0.02 \pm 0.018$  &  $0.02 \pm 0.054$ \\
        $k_{4}$   &  $15.2$      &  $12.8 \pm 1.251$  & $12.1 \pm 0.372$  &  $5.71 \pm 0.521$ \\
    \bottomrule
    \end{tabularx}
    \caption{List of the kinetic parameters that served as ground truth and KINN's predicted values for single-pulse ideal (Section \ref{subsec:results.1}), multi-pulse ideal (Section \ref{subsec:results.2}), and multi-pulse practical (Section \ref{subsec:results.3}) simulations with the standard deviations. Initial guesses of all parameters in all three cases were set to $1.0\times10^{-5}$.}
    \label{tab:pred_params}
\end{table*}

We note that the uncertainty estimates for parameters extracted from KINNs are based on the inverse Hessian. These estimates are not rigorous, since they assume that errors from the states and derivatives are uncorrelated with a constant variance of $\alpha$ \citep{Gusmao2023Maximum-likelihoodProblems}. Therefore, the uncertainty should be interpreted as a semi-quantitative measure of relative parameter sensitivity, rather than a quantitative error estimate, which explains the systematic under-estimation of errors in Table \ref{tab:pred_params}. A more rigorous error estimation for KINNs is beyond the scope of this work, and can be found in the work by \citet{Gusmao2023Maximum-likelihoodProblems}.

\subsection{Multi-pulse ideal scenario} 
\label{subsec:results.2}

In contrast to single-pulse experiments, where the catalyst state remains unchanged throughout the process, multi-pulse experiments gradually alter the surface catalyst state, leading to different outlet response and concentration profiles from pulse to pulse. In this case, we used a 10-pulse CO oxidation data set generated from 10 consecutive single-pulse experiments as a case study. The initial state of the coverage for each pulse corresponds to the final state of the preceding pulse. In real experimental scenarios, multi-pulse tests may involve hundreds or thousands of pulses, and may have unmeasured gaps between pulses. These factors pose significant computational challenges for data analysis since models must be solved sequentially and an uncontrolled error may be introduced due to gaps in the measurements. Hence, the ability to interpolate between multi-pulse data points and predict the initial concentrations of a given pulse becomes crucial for subsampling and analyzing dense TAP data.

To simulate altered states between pulses and evaluate the pulse subsampling capability, we divide the data set into a ``training'' set comprising pulses 0, 1, 2, 5, and 8, and a ``testing'' set comprising pulses 3, 4, 6, 7, and 9. Notably, the purpose of the testing set in this context is not to control for overfitting, as is common in the machine learning field, but rather to evaluate the ability of the model to represent the entire 10-pulse data set using 5 selected pulses.

\begin{figure*}[h]
    \centering\includegraphics[keepaspectratio=true,scale=0.45]{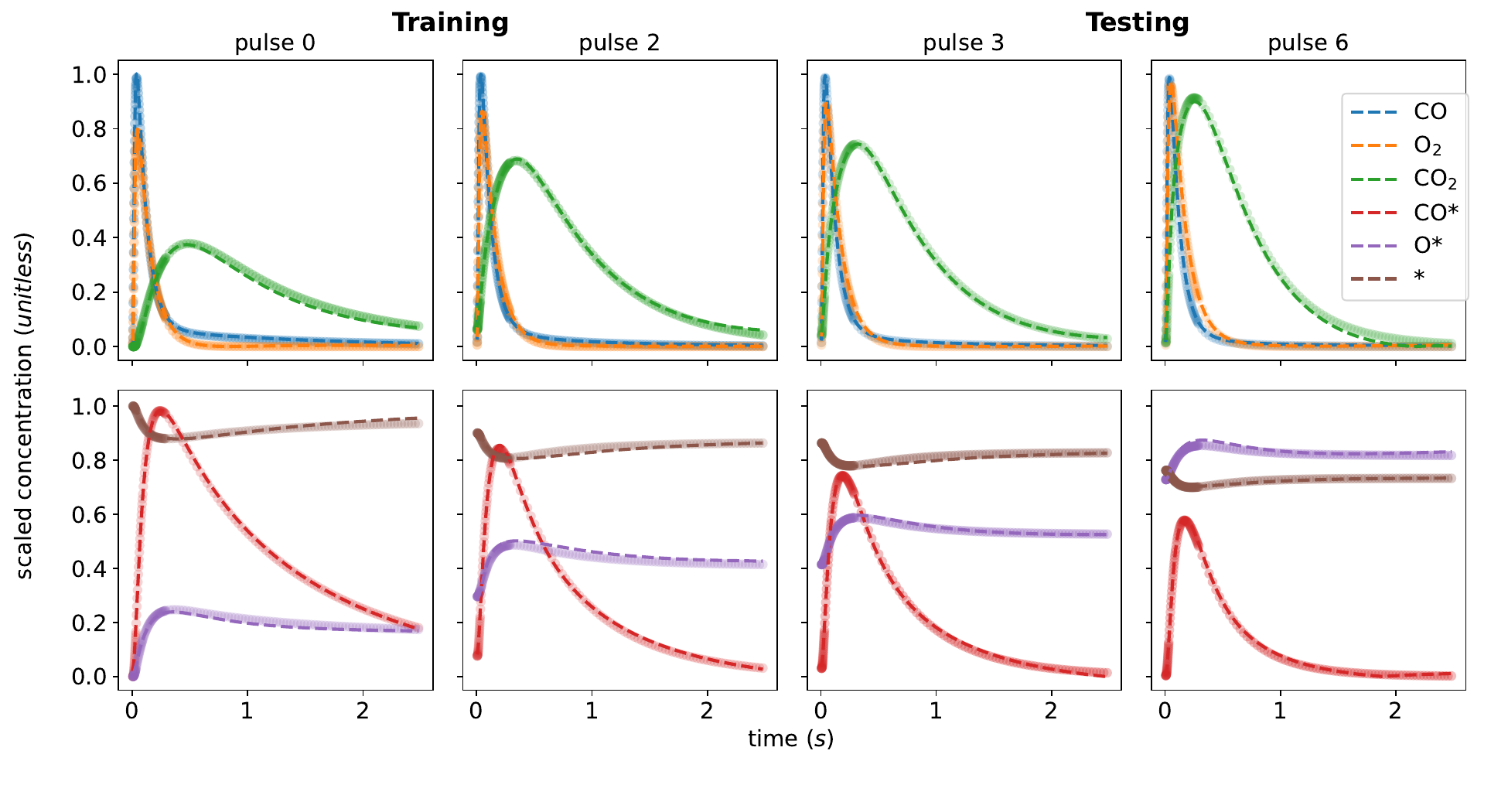}
    \caption{KINN's predicted scaled concentration (dashed line) and target ground truth value (dot) for the training set (pulse 0, 2) and interpolating to the testing set (pulse 3, 6) for gas species (top) and adspecies (bottom).}
    \label{fig:CO_multi_sub}
\end{figure*}

As shown in Fig. \ref{fig:CO_multi_sub}, the KINN is able to accurately fit the training data with an MAE of $6.29\times10^{-3}$ and accurately interpolate the testing data with a very similar MAE of $6.92\times10^{-3}$. In this case, we included more initial pulses in the training set to provide the NN with more data on kinetics at low coverages where the responses are more dynamic. As shown for the concentrations of the adspecies in Fig. \ref{fig:CO_multi_sub}, the KINN successfully captures the CO adsorption/desorption, the accumulation of adsorbed oxygen, and the reduction of available active sites. These results indicate that the KINN is able to accurately model the change in surface states over pulses.

The values of the kinetic parameters extracted from the ideal multi-pulse data are shown in Table \ref{tab:pred_params}. Compared to the results from the single pulse data, the MAE of the extracted kinetic parameters, in the free energy of activation scale, has decreased from 0.019 eV to 0.016 eV. This reduction is also shown in Fig. \ref{fig:parity_compare}. Although the improvement in model precision is modest, additional testing also shows that inclusion of multiple pulses systematically improves the accuracy of the kinetic parameters (see the Supplementary Information). Moreover, the decrease in the standard deviation of $k_4$ indicates that multi-pulse fitting yields a more robust estimate of this parameter compared to single-pulse fitting. This improved resolution of this parameter in the case of multi-pulse data is likely due to the Eley-Rideal $CO_2$ formation mechanism, where $CO$ reacts directly with adsorbed oxygen. This reaction is expected to be more sensitive to ``state-altering'' experiments where the O* coverage increases, increasing the rate and contribution of this particular elementary step \citep{Persson1995FlatSurface, Zhu2023InvestigatingSurface}. This result is consistent with the intuition that the multi-pulse data will be sensitive to more kinetic parameters, highlighting the particular importance of multi-pulse fitting for capturing the nuances of reactions involving adsorbed species.

\subsection{Multi-pulse practical scenario}
\label{subsec:results.3}

In real experiments, measured data contains noise, and it is not possible to directly measure the catalyst zone concentrations and net fluxes. To address this, as mentioned in Section \ref{subsec:methods.2}, the Y-Procedure was introduced to estimate the thin zone gas phase concentrations, net fluxes, and atomic uptake. The inclusion of atomic uptake, which restricts the concentrations of unobservable adspecies, requires the modification of the loss function $J$ into Eq. \ref{eqn:loss_fn_uptake},
\begin{equation} \label{eqn:loss_fn_uptake}
    \min_{\boldsymbol{\omega}} \quad J(\boldsymbol{\omega}) = j_{data}(\boldsymbol{\omega_s}) + \alpha j_{model}(\boldsymbol{\omega_s}, \mathbf{k}) + \beta j_{uptake}(\boldsymbol{\omega_s})
\end{equation}
where $\beta$ is the weighted parameter for uptake loss $j_{uptake}$, and is found in the similar way as $\alpha$. 

\begin{figure*}[h]
    \centering
    \includegraphics[keepaspectratio=true,scale=0.5]{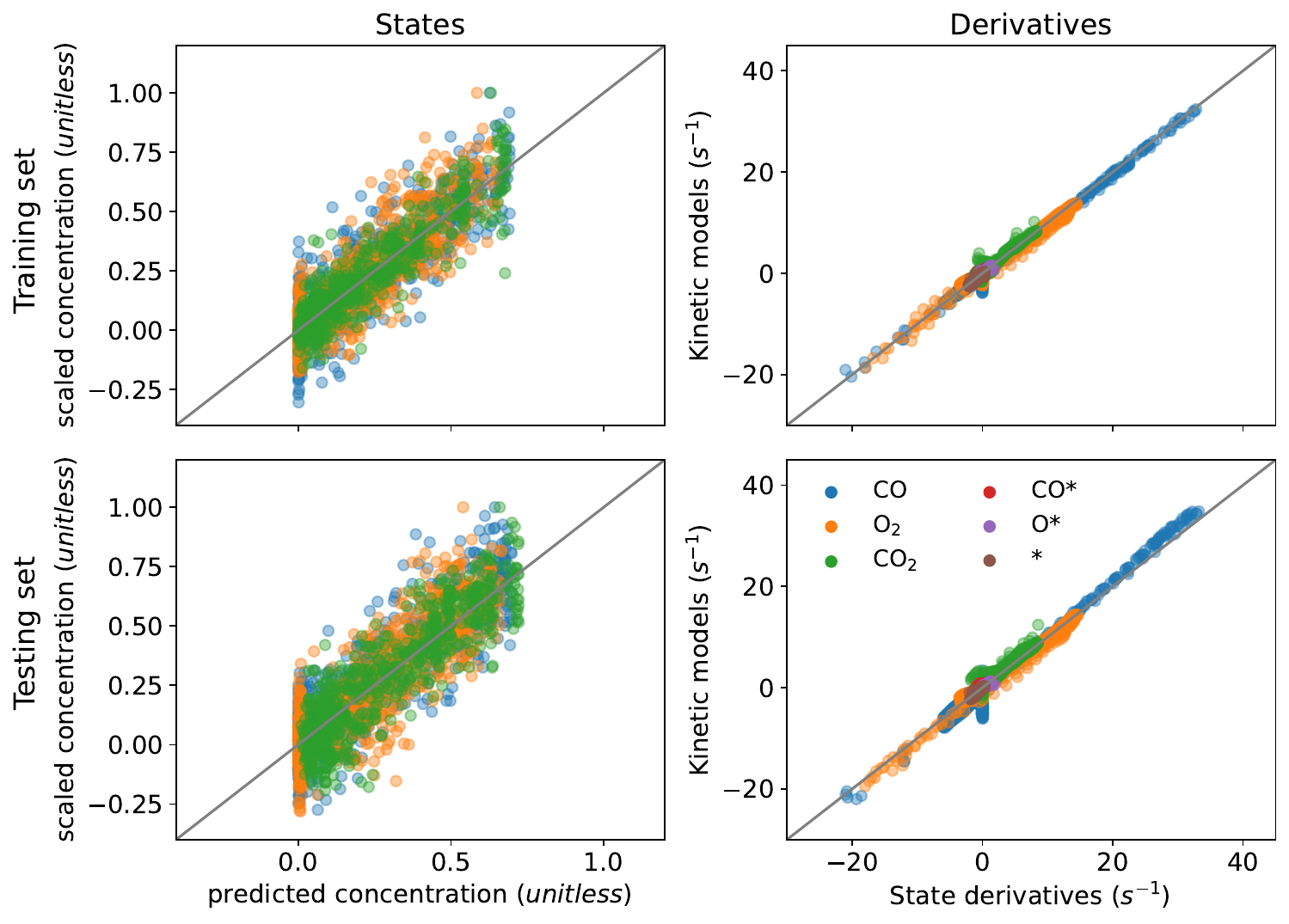}
    \caption{Training (top) and testing (bottom) performance for practical multi-pulse case under minimization of Eq.\ref{eqn:loss_fn_uptake}. }
    \label{fig:parity_compare}
\end{figure*}

Application of the KINN method to the practical multi-pulse data yields the parity plots shown in Fig. \ref{fig:parity_compare}. These plots reveal good agreement between the KINN predicted concentrations and kinetics and the target values for training and testing pulses. The results show significant deviation in the concentration fitting due to the noise, but reveal strong agreement for the model fitting. Despite the fact that the concentrations of adspecies are not available through direct measurement, the additional $j_{uptake}$ term grants the network knowledge about the surface environment in the form of the total amount of each element that remains on the surface. As a result, penalizing this term with proper weighting parameters can limit the concentrations of adspecies to physically meaningful values, at least in the case of a simple reaction network. These results also highlight the robustness of KINNs to noise. The high correlations in Fig. \ref{fig:parity_compare} indicate that the model can accurately predict both concentrations and derivatives despite the presence of noise. This robustness is a key feature of KINNs that results from simultaneous optimization of the interpolating function and the underlying model in a single loss function. We note that as the reaction complexity increases, the uptake constraint is expected to become less effective. Nevertheless, the results still show that it can be used as an effective soft constraint in the optimization problem. It is straightforward to envision that the loss function can be extended to information from different sources, such as thermodynamic constraints or spectroscopic signals, to provide constraints on adspecies concentrations in more complicated models.

Table \ref{tab:pred_params} demonstrates that all kinetic parameters were solved within the correct order of magnitude with an energy MAE of $0.034$ eV in this case, although the fitted values show a greater deviation from the ground truth compared to the ideal cases, as shown in Fig. \ref{fig:parity_compare}. This deviation is expected due to the introduction of noise and the substitution of precise thin zone information with estimated quantities. Consequently, the larger error bars suggest that the KINN exhibits reduced sensitivity to all parameters in the presence of noise. Moreover, the inherent information loss in estimating concentrations and fluxes also contributes to performance degradation. The subsequent analysis indicates that KINN parameter estimates are robust to noise, suggesting that the primary cause of the reduced accuracy is the lack of direct information about the adsorbate concentrations.
Despite the reduced accuracy in kinetic parameters compared to ideal cases, the results still provide parameters with energy errors well below what is common in DFT, and the accuracy could be further improved by using the KINN results as initial guesses for more rigorous and computationally demanding PDE-based TAP analysis tools, such as TAPSolver\citep{Yonge2021TAPsolver:Experiments}.

\begin{figure*}[h]
    \centering
    \includegraphics[keepaspectratio=true,scale=0.45]{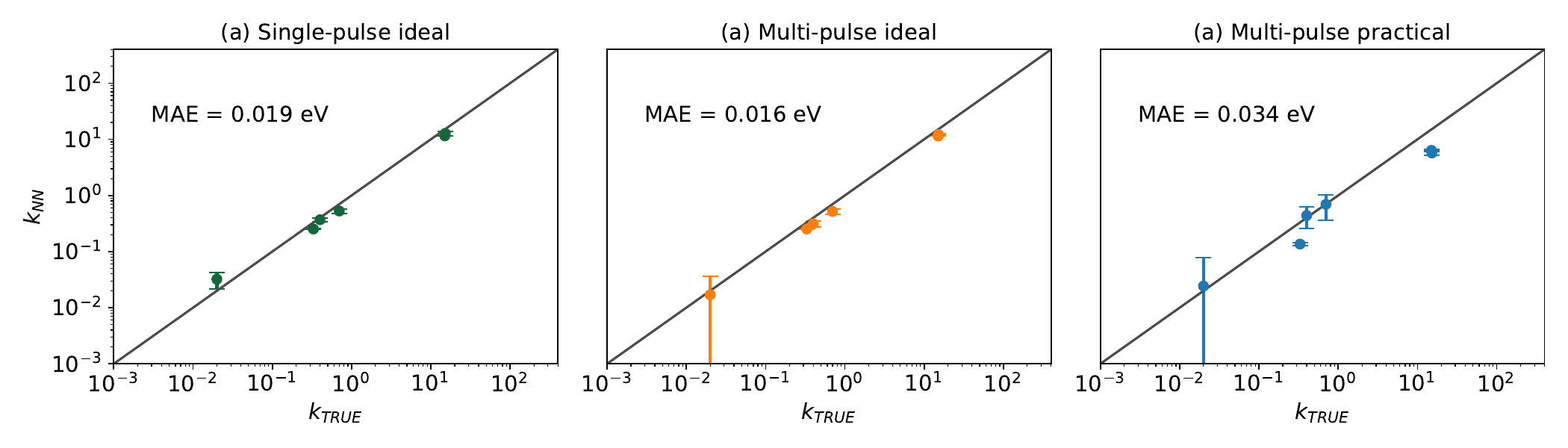}
    \caption{Logarithm parity between the ground truth and KINNs solved kinetic parameter values that presented in Table \ref{tab:pred_params}. The MAE is calculated using the Eyring equation.}
    \label{fig:uptake_kinetics}
\end{figure*}

\subsection{Multi-pulse analysis compared to differential algebraic programming}
\label{subsec:comp_pyomo}

To highlight the advantages of the KINN approach, we directly compare its performance with an established optimization approach based on differential algebraic equations (DAEs) and collocation. This comparison is performed with Pyomo, a Python-based open-source optimization modeling tool \citep{Hart2011Pyomo:Python, Bynum2021PyomoPython, Nicholson2018Pyomo.dae:Equations, Biegler2010NonlinearProcesses}. The ideal multi-pulse data set was used for the comparison, and noise was added in the same way as described in Section \ref{subsec:methods.3}. In addition to Gaussian noise with a width of half the data standard deviation, we explored two other noise levels: one and two standard deviations. In this case, KINN was applied directly to unsmoothed noisy data, while the DAE collocation approach was tested on both unsmoothed data and data smoothed by the Savitzky-Golay filter \citep{Savitzky1964SmoothingProcedures}. The Pyomo script and the detailed approach can be found in the Supplementary Information.

Fig. \ref{fig:kinn_py_forward} indicates that the KINN approach has improved noise tolerance. As the noise level increases, the concentration fitting of both smoothed and unsmoothed DAE collocation starts to deviate from the ground truth and exhibits discontinuous noisy behavior. The collocation method is much faster, by a factor of $\sim$100 with the implementation and architectures tested here, although we note that the KINNs approach is not heavily optimized and run times will depend on many factors. However, the collocation approach shows reduced effectiveness at higher noise levels. This is because the discretization of the time domain used for approximating solutions in the collocation method may not adequately capture the random fluctuations caused by noise, resulting in an inaccurate approximation of the true solution, and hence an inaccurate estimation of the derivatives. Moreover, collocation methods typically rely on interpolation and smoothing techniques to estimate the dynamics of the system between discretized points, and these
techniques can be sensitive to noise and numerical parameters \citep{Craven1978SmoothingCross-validation, Knowles2012MethodsData}.At high noise levels, it also becomes impossible to converge estimates for some parameters with the collocation technique. In contrast, the KINN approach converges all parameters across different noise levels. This is evident from the lower deviation of the KINN-predicted concentrations compared to the DAE results in the presence of noise. The neural network in the KINN acts as a highly flexible built-in interpolator, simultaneously solving the smoothing and fitting problems. Moreover, the optimization of the KINN is constrained and regularized by the kinetic model, which helps prevent overfitting to the noise. This robustness to noise is a key advantage of the KINN over traditional collocation methods, as it ensures a more reliable approximation of the underlying kinetics even when noise levels are high or datasets have missing values.

\begin{figure*}[h]
    \centering
    \includegraphics[keepaspectratio=true,scale=0.55]{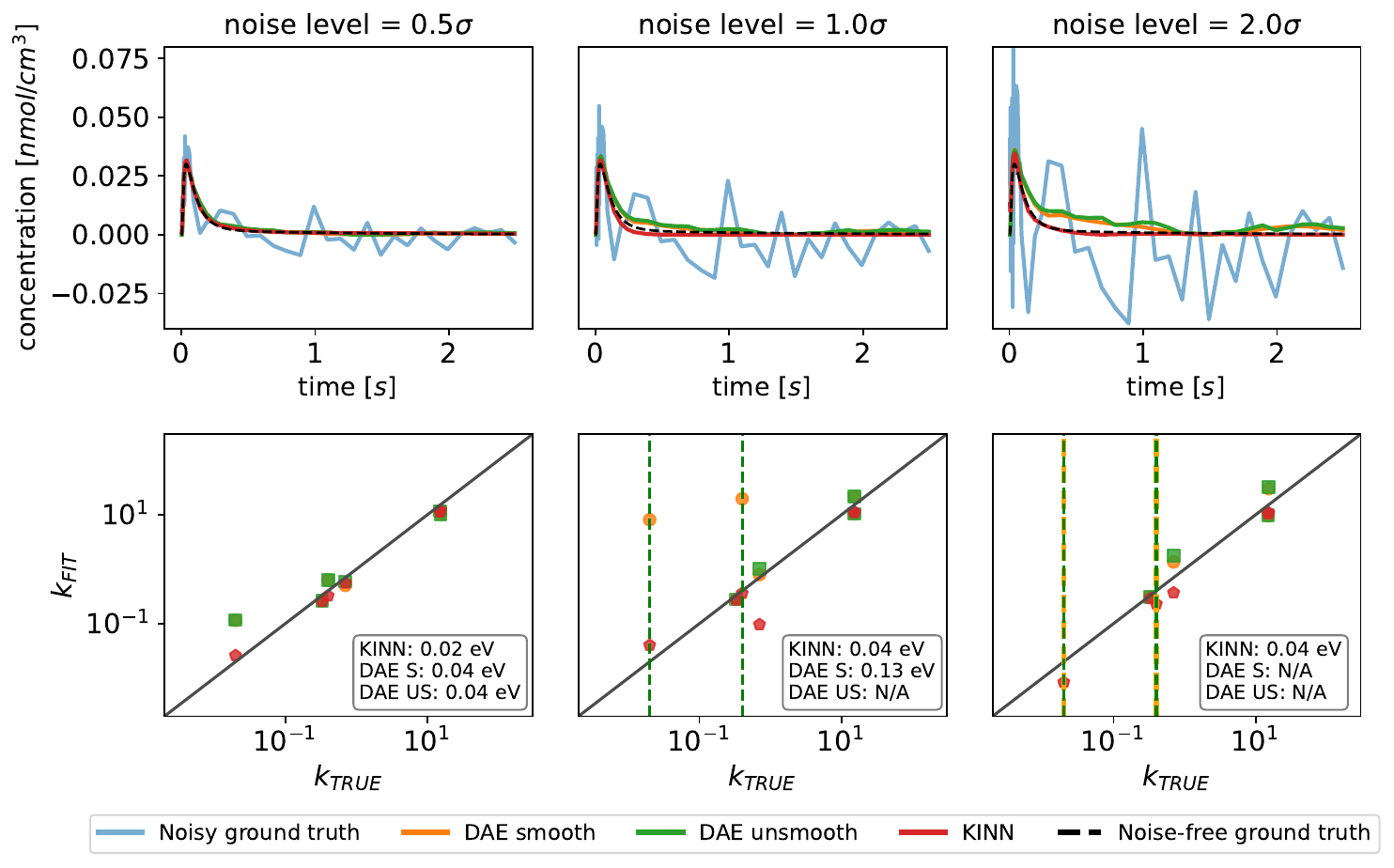}
    \caption{Performance of DAE and the KINN on fitting data with increasing noise level. The vertical dash lines in kinetic parameter parity plots means the specific parameters are not converged. CO in pulse 0 is used as the example species here. The MAEs in an energy scale are stated in parity plots. The complete multi-pulse concentration plots can be found in the Supplementary Information.}
    \label{fig:kinn_py_forward}
\end{figure*}

For the extracted kinetic parameters, the KINN also demonstrates better performance than DAE collocation for this case study. As shown in Fig. \ref{fig:kinn_py_forward}, at a noise level of $0.5\sigma$, the MAE for free energy of activation extracted by the KINN is 0.018 eV, only slightly higher than the 0.016 eV MAE for the ideal multi-pulse scenario without noise. This outperforms the DAE collocation approach which has an MAE of 0.04 eV regardless of whether or not smoothing is applied. When the noise level increases to $1.0\sigma$, the unsmoothed DAE fails to converge for two parameters, and at a noise level of $2\sigma$, the convergence issue extends to both smoothed and unsmoothed DAE collocation, while the KINN continues to deliver consistent and accurate results comparable to the $0.5\sigma$ noise scenario. The values of the fitted parameters for all reactions with each method can be found in the Supplementary Information. 

There are several factors that account for this notable improvement in robustness. Primarily, the KINN provides a more accurate concentration fitting. Because the kinetics depend on these concentration values, this accuracy directly enhances the quality of the extracted parameters. Moreover, traditional collocation methods compute derivatives using exact data points, which can amplify the impact of noise on the estimation of kinetic parameters. In contrast, KINNs utilize automatic differentiation during backward propagation. As described in Section \ref{subsec:methods.3}, this differentiation is applied to an interpolated concentration profile instead of the noisy data. This difference is crucial because it allows KINNs to derive kinetic parameters that are more resilient to noise, resulting in closer approximation to the ground truth in high-noise scenarios.

\section{Conclusions}
\label{sec:conc}

In this work, we utilized feedforward NNs as basis functions to describe  TAP data and extract the intrinsic kinetics from the approximated ODE MKMs under state-defining (single-pulse) and state-altering (multi-pulse) CO oxidation data sets. The validity of KINNs is demonstrated by their ability to solve the underlying ODEs, extract the rate parameters of kinetic models, and forecast the responses of unobserved pulses in both scenarios. Applying zeroth moments as an additional NN input provides the network with information on the surface environment in multi-pulse datasets and enables the network to locate the pulse in pulse series, making interpolation between pulses achievable. 

The machine learning framework grants KINNs scalability to handle multi-pulse data with a single network, and tests indicate that KINNs exhibit improved noise tolerance when compared to the incumbent DAE collocation approach. With these features, dense, noisy data consisting of hundreds or thousands of pulses can be subsampled and analyzed, revealing underlying kinetics that cannot be inspected through state-defining experiments only. Currently, the main downside of KINNs is the relatively high computational cost, but with further optimization and increasing problem complexity, their advantages over the DAE approach are expected to increase. For example, the KINN framework is naturally capable of handling highly nonlinear problems such as temperature-dependent data or coverage-dependent kinetic parameters. Future work will focus on exploring the application of KINNs to these more complex data types, optimizing the KINN implementation, and incorporating the maximum likelihood approach \citep{Gusmao2023Maximum-likelihoodProblems} for data where not all concentrations are known .

The flexible loss function makes KINNs a solid option for addressing complex heterogeneous catalytic reactions. We show that when the precise thin zone information is unavailable, KINNs can still derive kinetic parameters that accurately reflect the relative rate of each step by including atomic uptake into the loss function. This flexibility allows for incorporating a variety of information, including spectroscopy, DFT, and thermodynamics, which may be used to constrain more complex systems than CO oxidation, suggesting KINNs as a promising framework for treatment of operando and in situ transient kinetic data sets. 

KINNs offer a robust new approach to analyzing kinetic models from TAP data. The use of neural networks as a basis set offers improved noise tolerance and the possibility to leverage computational tools from the machine learning field. Furthermore, the physically interpretable outputs (rate parameters) make the results obtained from KINNs readily available for further evaluation and application in kinetic simulations and process models. Future work will focus on improved strategies for identifying hyperparameters of KINNs, improving computational scalability, and applying them to more complex transient kinetic datasets. We expect that KINNs will lower the multi-pulse TAP data analysis barrier and provide insight into fusing contemporary computational methods with transient catalytic dynamics.

\section*{Acknowledgements}
This work was supported by the U.S. Department of Energy (USDOE), Office of Energy Efficiency and Renewable Energy (EERE), Industrial Efficiency and Decarbonization Office (IEDO), Next Generation R\&D Project DE-FOA-0002252-1175 under contract no. DE-AC07-05ID14517. This research was supported in part through research cyberinfrastructure resources and services provided by the Partnership for an Advanced Computing Environment (PACE) at the Georgia Institute of Technology, Atlanta, Georgia, USA. This research also made use of Idaho National Laboratory’s High Performance Computing systems located at the Collaborative Computing Center supported by the Office of Nuclear Energy of the U.S. Department of Energy and the Nuclear Science User Facilities under Contract No. DE-AC07-05ID14517. The authors are grateful to Dr. Adam Yonge for the discussion of TAP simulations and transient methods and to Prof. Joseph Scott for suggestions on comparing to existing methods.

\bibliography{main}
\bibliographystyle{plainnat} 

\end{document}


\maketitle

\let\thefootnote\relax\footnotetext{$*$~Corresponding author. Email: ajm@gatech.edu}

\section{TAPSolver Setup}
\label{sec:sup.1}

A sample TAPSolver script we used to generate the synthetic CO oxidation data is provided. The $forward\_problem(pulse\_time, pulse\_number, TAPobject)$ function can be adjusted for different pulse duration or performing a multi-pulse simulation.

\begin{lstlisting}[language = Python]
import json

from tapsolver import *

new_reactor = reactor()

new_reactor.zone_lengths = {0: 2.80718, 1: 0.17364, 2: 2.80718}  # cm
new_reactor.zone_voids = {0: 0.4, 1: 0.4, 2: 0.4}  # -
new_reactor.reactor_radius = 1  # cm

new_reactor_species = reactor_species()
new_reactor_species.inert_diffusion = 16  # cm2/s
new_reactor_species.catalyst_diffusion = 16  # cm2/s
new_reactor_species.reference_temperature = 385.5  # K
new_reactor_species.reference_mass = 60  # 24.01 # amu
new_reactor_species.temperature = 385.65  # K
new_reactor_species.advection = 0

CO = define_gas()
CO.mass = 28.01
CO.intensity = 1
CO.delay = 0
CO.noise = 0.0
new_reactor_species.add_gas('CO', CO)

O2 = define_gas()
O2.mass = 32
O2.intensity = 1
O2.delay = 0.0
O2.noise = 0.0
new_reactor_species.add_gas('O2', O2)

CO2 = define_gas()
CO2.mass = 44.01
CO2.intensity = 0
CO2.delay = 0.0
CO2.noise = 0.0
new_reactor_species.add_gas('CO2', CO2)

Ar = define_gas()
Ar.mass = 40
Ar.intensity = 1
Ar.delay = 0.0
Ar.noise = 0.0
new_reactor_species.add_inert_gas('Ar', Ar)

s = define_adspecies()
s.concentration = 0
new_reactor_species.add_adspecies('CO*', s)

s = define_adspecies()
s.concentration = 0
new_reactor_species.add_adspecies('O*', s)

s = define_adspecies()
s.concentration = 30
new_reactor_species.add_adspecies('*', s)

new_mechanism = mechanism()

new_mechanism.elementary_processes[0] = elementary_process('CO + * <-> CO*')
new_mechanism.elementary_processes[1] = elementary_process('O2 + 2* -> 2O*')
new_mechanism.elementary_processes[2] = elementary_process('CO* + O* <-> CO2 + 2*')
new_mechanism.elementary_processes[3] = elementary_process('CO + O* -> CO2 + *')

new_mechanism.elementary_processes[0].forward.k = 15
new_mechanism.elementary_processes[0].backward.k = 0.7
new_mechanism.elementary_processes[1].forward.k = 0.33
new_mechanism.elementary_processes[1].backward.k = 0
new_mechanism.elementary_processes[2].forward.k = 0.4
new_mechanism.elementary_processes[2].backward.k = 0.02
new_mechanism.elementary_processes[3].forward.k = 15.2
new_mechanism.elementary_processes[3].backward.k = 0


#Make a function
for j in new_mechanism.elementary_processes:
    new_mechanism.elementary_processes[j].forward.use = 'k'
    try:
        new_mechanism.elementary_processes[j].backward.use = 'k'
    except:
        pass

mechanism_constructor(new_mechanism)

CO_tapobject = TAPobject()
CO_tapobject.store_flux_data = True
CO_tapobject.store_thin_data = True
CO_tapobject.store_cat_flux = True
CO_tapobject.display_analytical = True
CO_tapobject.show_graph = True
CO_tapobject.mechanism = new_mechanism
CO_tapobject.reactor_species = new_reactor_species
CO_tapobject.reactor = new_reactor

# CO_tapobject.show_graph = False
CO_tapobject.output_name = 'CO_TAP_long_inert'

CO_tapobject.gasses_objective = ['CO', 'O2', 'CO2']  # ['Ar']
CO_tapobject.optimize = False

forward_problem(2.5, 1, CO_tapobject)
flux_graph(CO_tapobject)
\end{lstlisting}

\section{Multi-pulse ideal scenario}

\begin{figure}[h]
    \centering
    \includegraphics[keepaspectratio=true,scale=0.5]{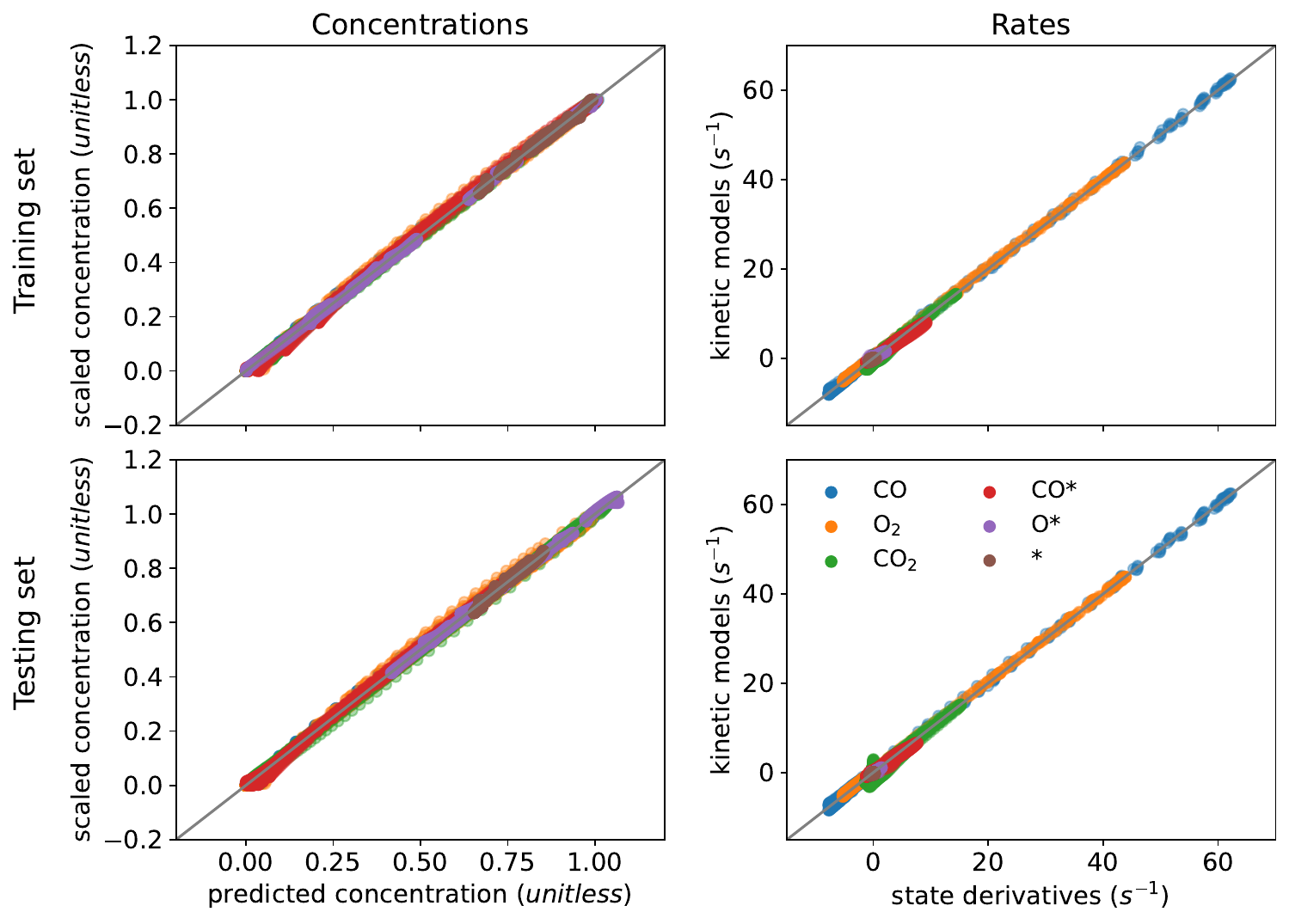}
    \caption{Training (top) and testing (bottom) performance on the concentration prediction and kinetic model fitting for multi-pulse ideal scenario.}
    \label{fig:parity_multi_ideal}
\end{figure}

\begin{figure}[H]
    \centering\includegraphics[keepaspectratio=true,scale=0.6]{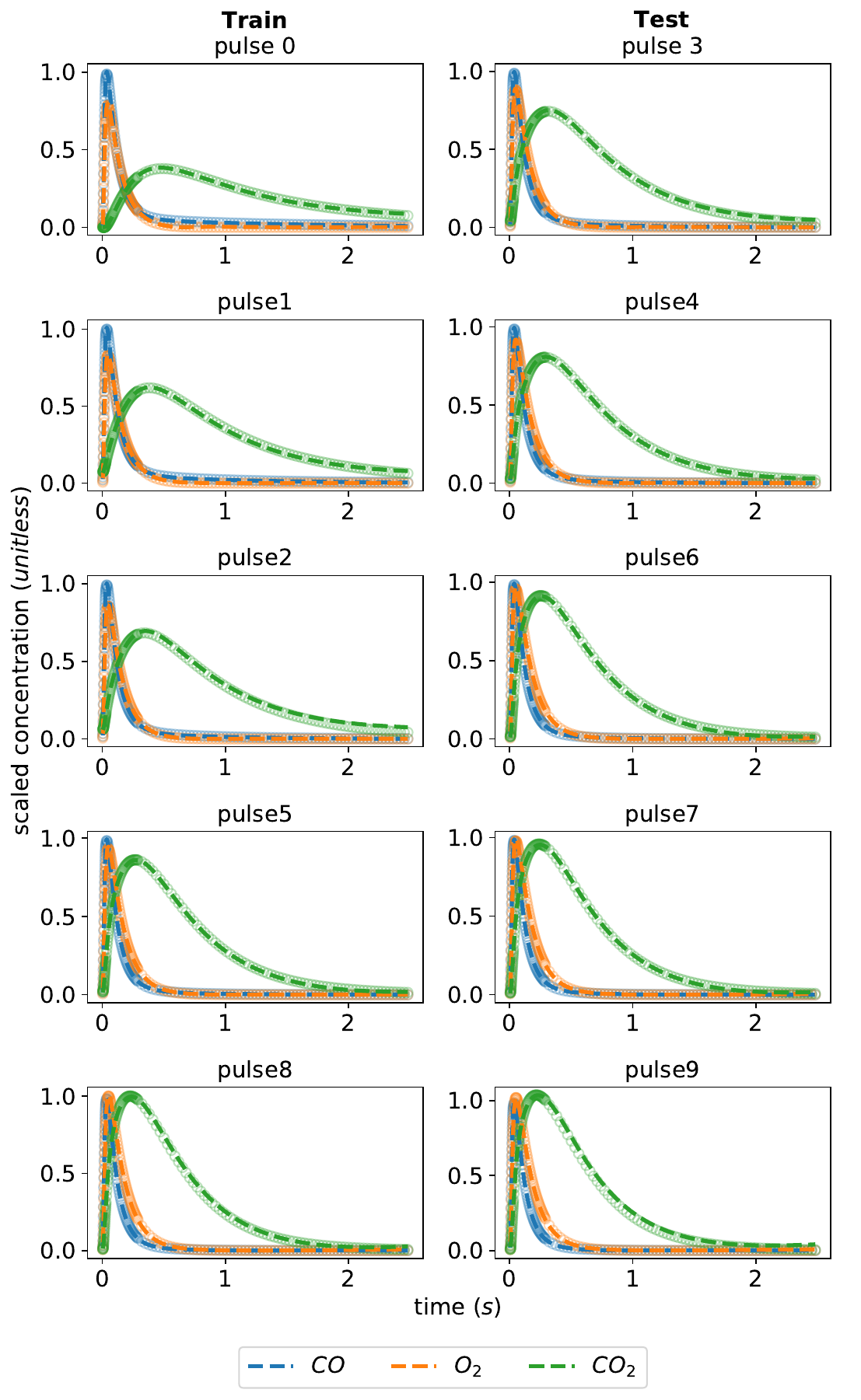}
    \caption{KINN's predicted scaled concentration (dash line) and target ground truth value (dot) for the training set and interpolating to the testing set for gas species.}
    \label{fig:multi_ideal_gas}
\end{figure}

\begin{figure}[H]
    \centering\includegraphics[keepaspectratio=true,scale=0.6]{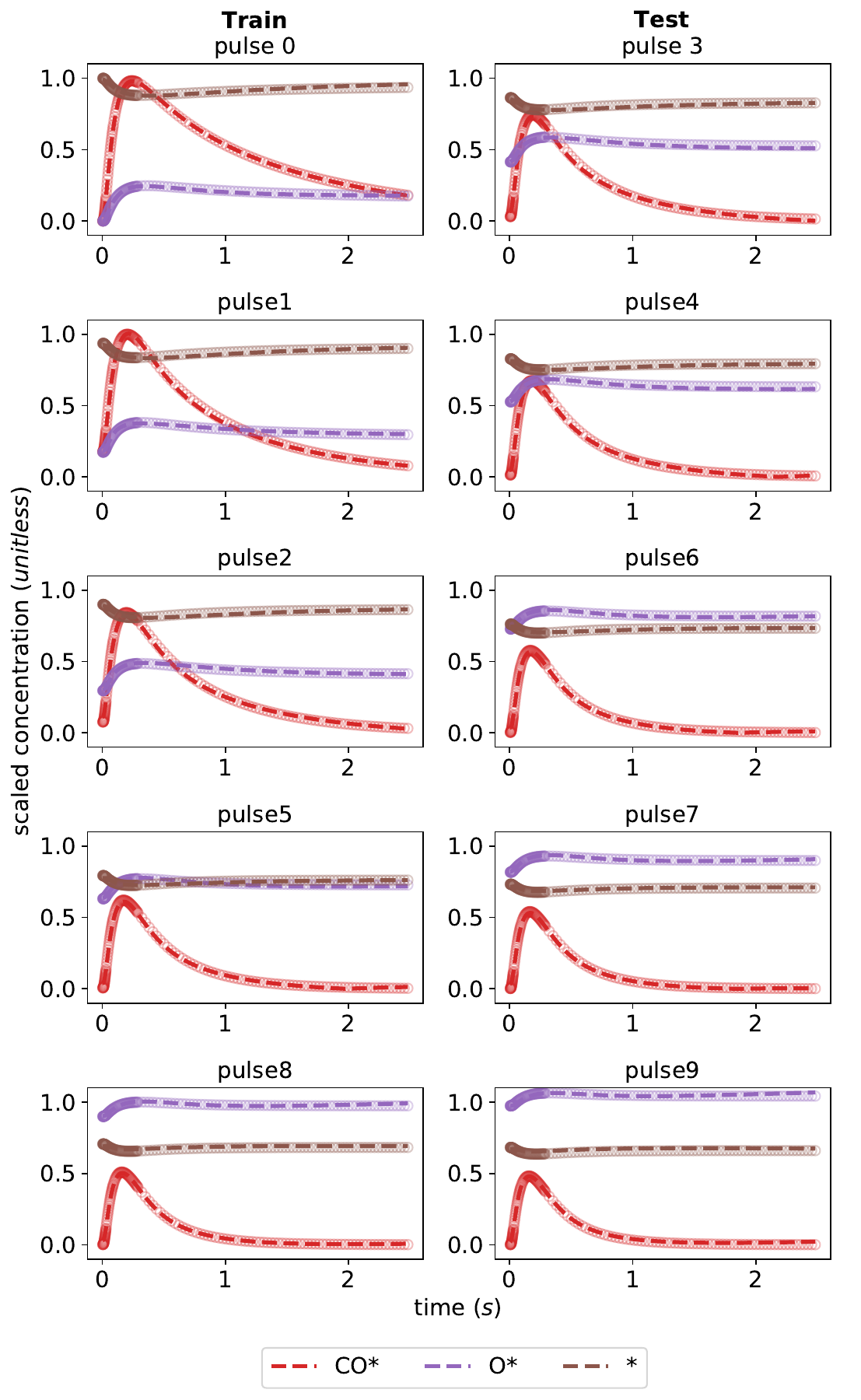}
    \caption{KINN's predicted scaled concentration (dash line) and target ground truth value (dot) for the training set and interpolating to the testing set for adspecies.}
    \label{fig:multi_ideal_ad}
\end{figure}

\section{Multi-pulse practical scenario}

\begin{figure}[H]
    \centering\includegraphics[keepaspectratio=true,scale=0.65]{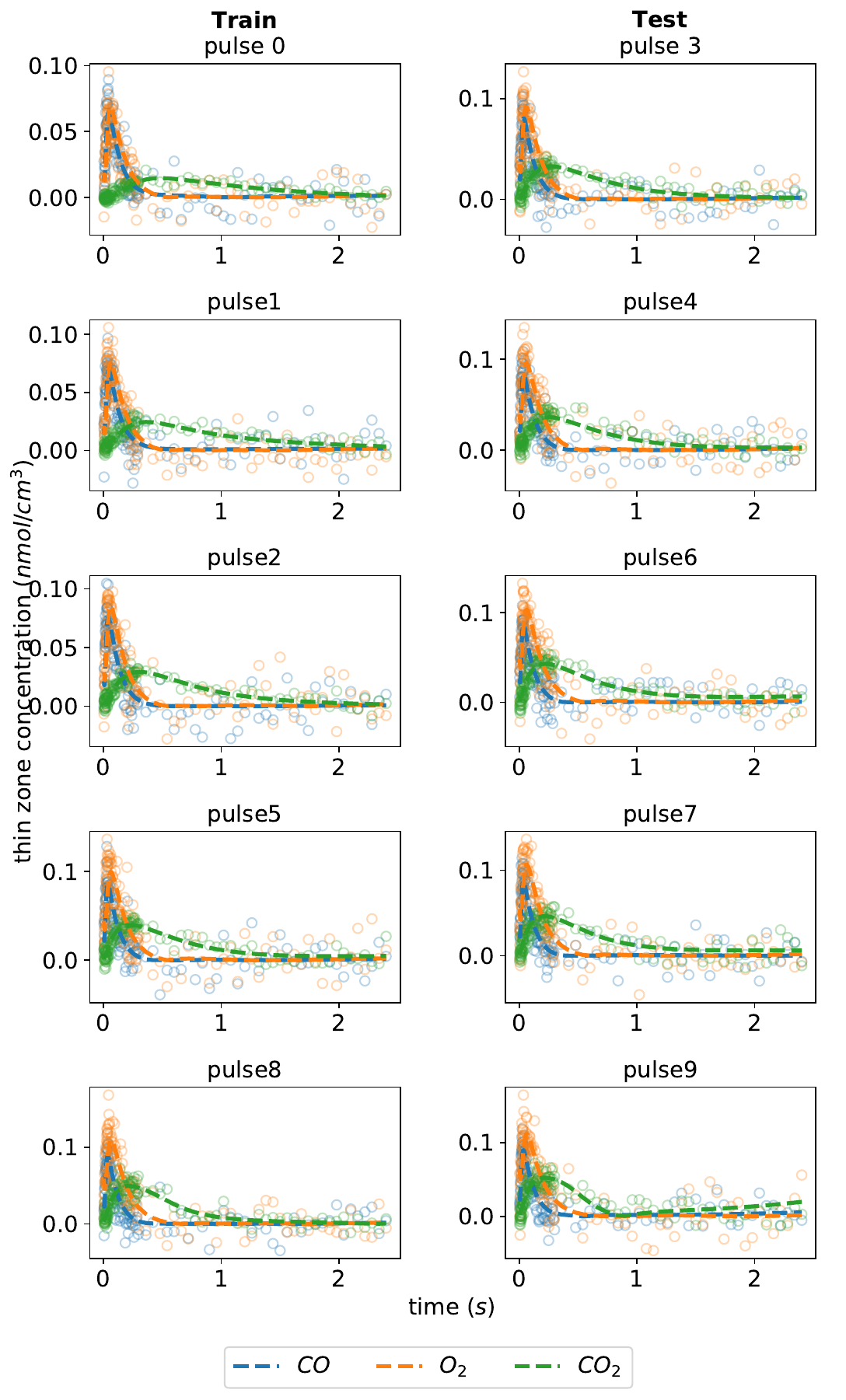}
    \caption{KINN's predicted scaled concentration (dash line) and target ground truth value (dot) for the training set and interpolating to the testing set for gas species. The discontinuity at 0 happens because the absolute value is taken at the output layer as we consider negative concentration is invalid, but the continuous states goes negative because of the noise. A use of softplus/swish layer can solve this discontinuity by test.}
    \label{fig:multi_prac_gas}
\end{figure}

\begin{figure}[H]
    \centering\includegraphics[keepaspectratio=true,scale=0.65]{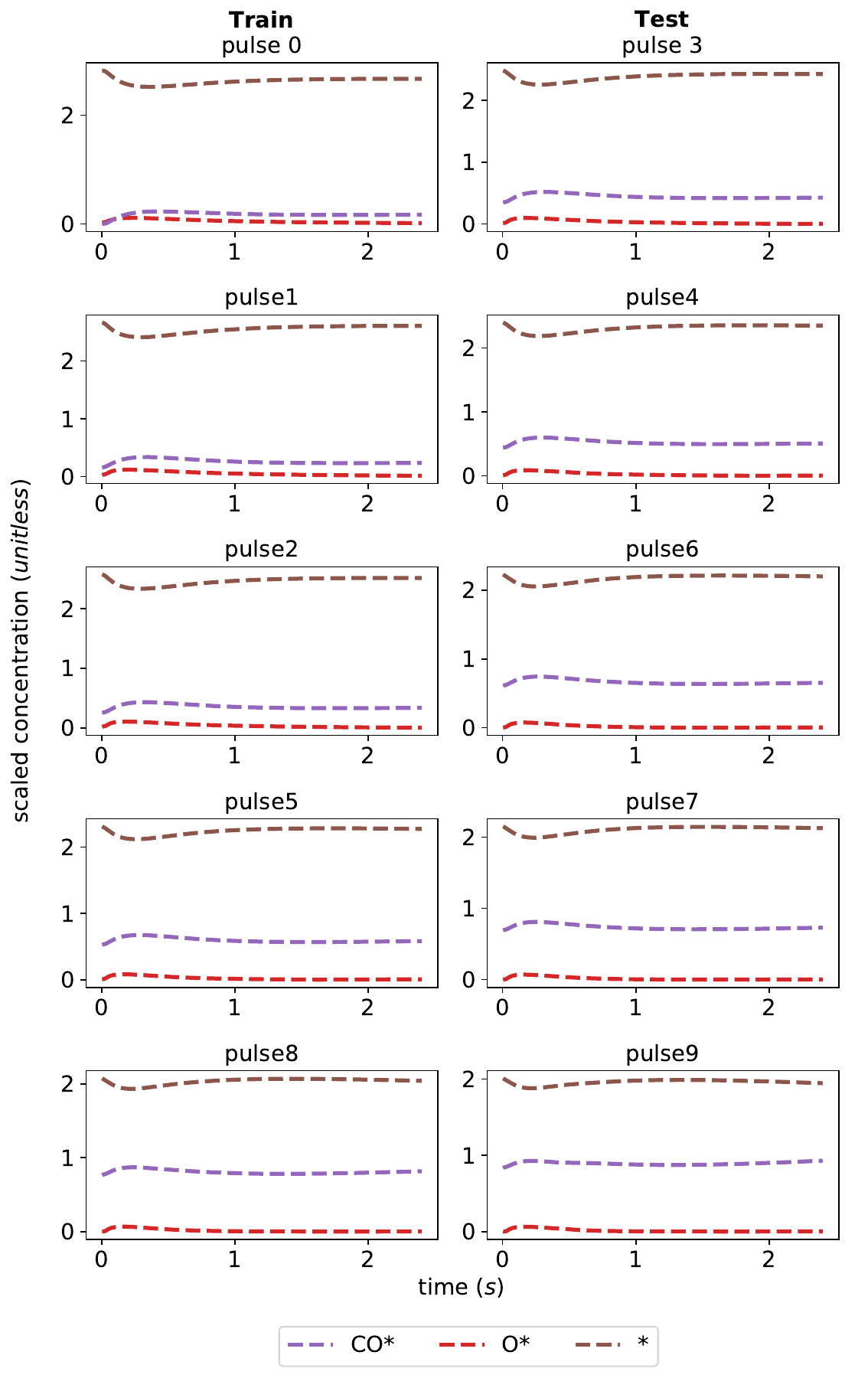}
    \caption{KINN's predicted scaled concentration (dash line) for the training set and interpolating to the testing set for adspecies. The ground truth concentrations are not available in practical scenario, so only the predicted values are shown. Compared to Figure \ref{fig:multi_ideal_ad}, the scaled concentrations are different because they are scaled with different values. In the ideal case, the adspecies concentrations are rescaled with the maximum ground truth values of each species; in the practical case, they are rescaled with the maximum atomic uptake value.}
    \label{fig:multi_prac_ad}
\end{figure}

\subsection{Multi-pulse training effect on kinetic model quality}


\begin{figure}[H]
    \centering\includegraphics[keepaspectratio=true,scale=0.6]{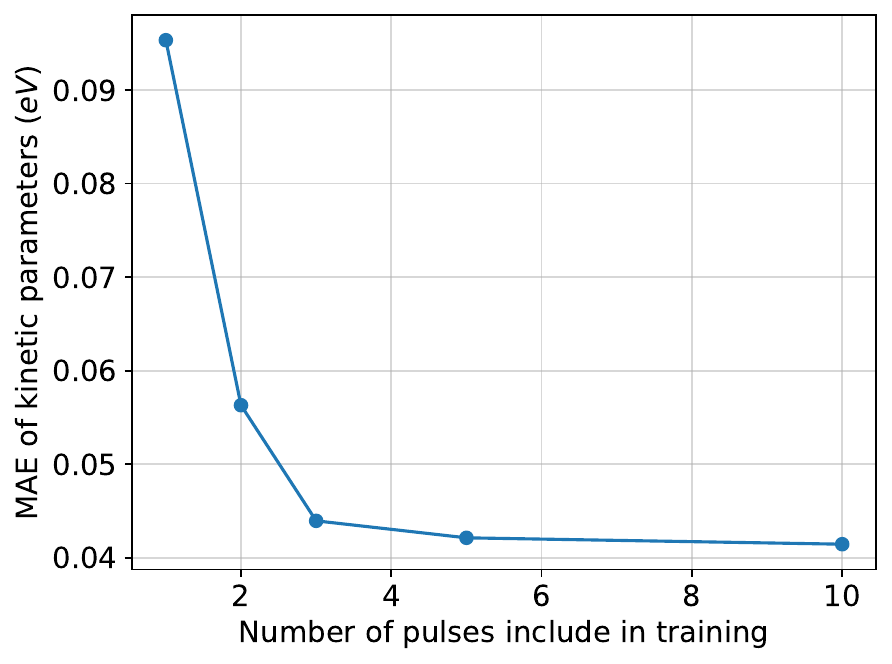}
    \caption{MAE of kinetic parameters as a function of the number of pulses included in training for the practical multi-pulse scenario. Note that the MAE is higher than reported in the main text since fewer epochs were used and the $\alpha$ value was not as carefully optimized.}
    \label{fig:pulse_mae}
\end{figure}

\section{Multi-pulse analysis compared to differential algebraic programming}

\begin{figure}[H]
    \centering\includegraphics[keepaspectratio=true,scale=0.5]{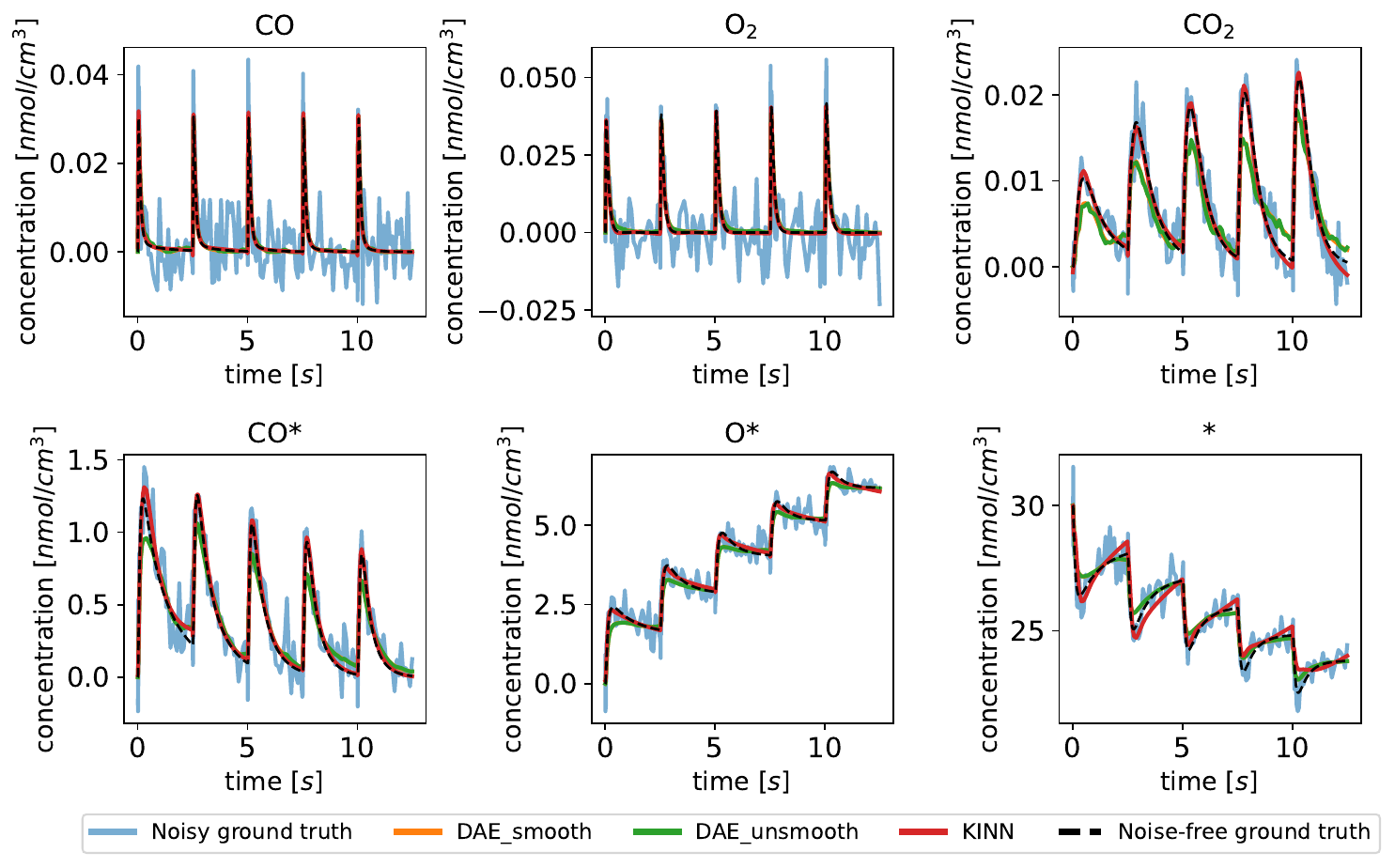}
    \caption{Concentration fitting of DAE and the KINN at 0.5$\sigma$ noise level.}
    \label{fig:dae_05}
\end{figure}

\begin{figure}[H]
    \centering\includegraphics[keepaspectratio=true,scale=0.5]{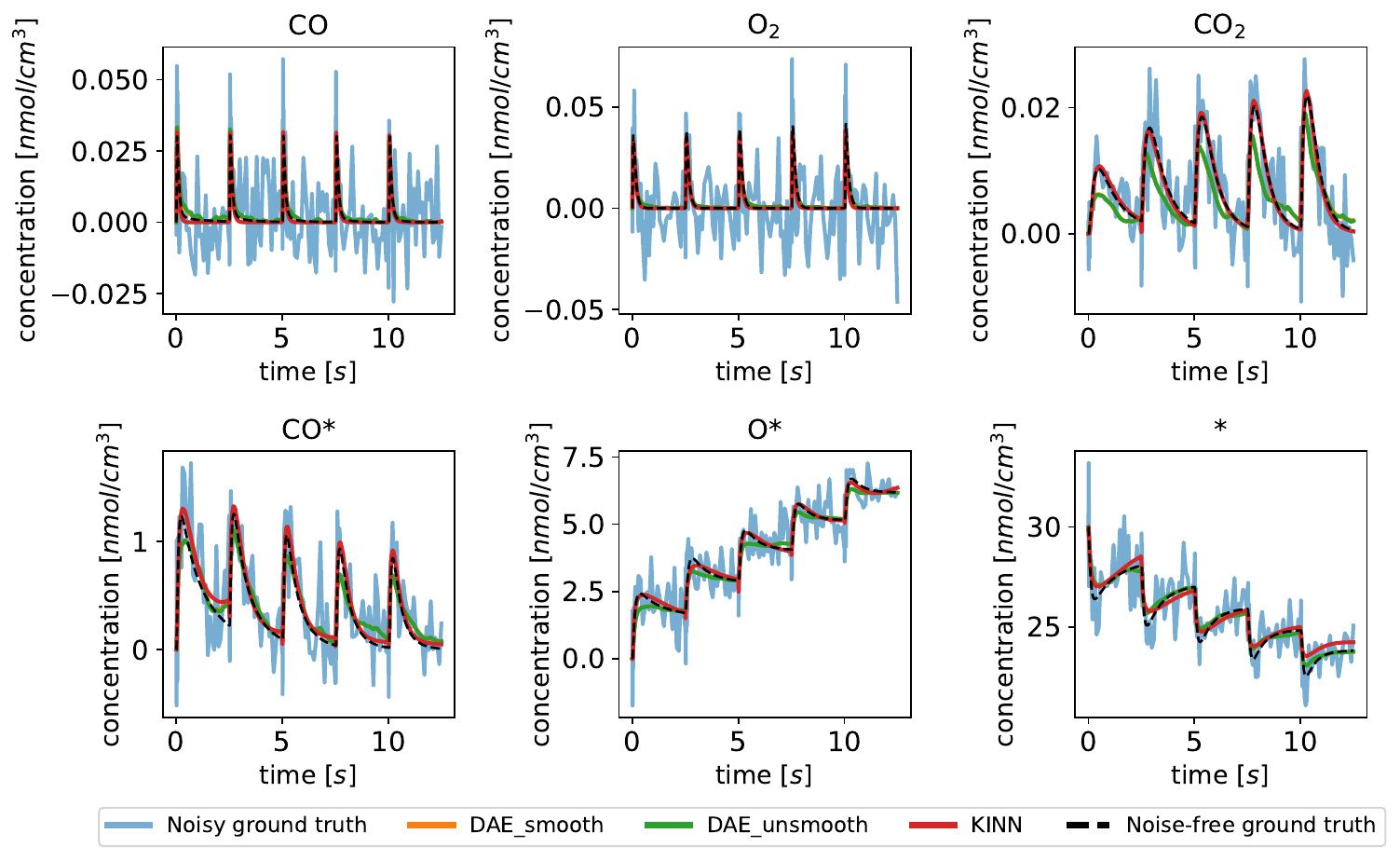}
    \caption{Concentration fitting of DAE and the KINN at 1.0$\sigma$ noise level.}
    \label{fig:dae_10}
\end{figure}

\begin{figure}[H]
    \centering\includegraphics[keepaspectratio=true,scale=0.5]{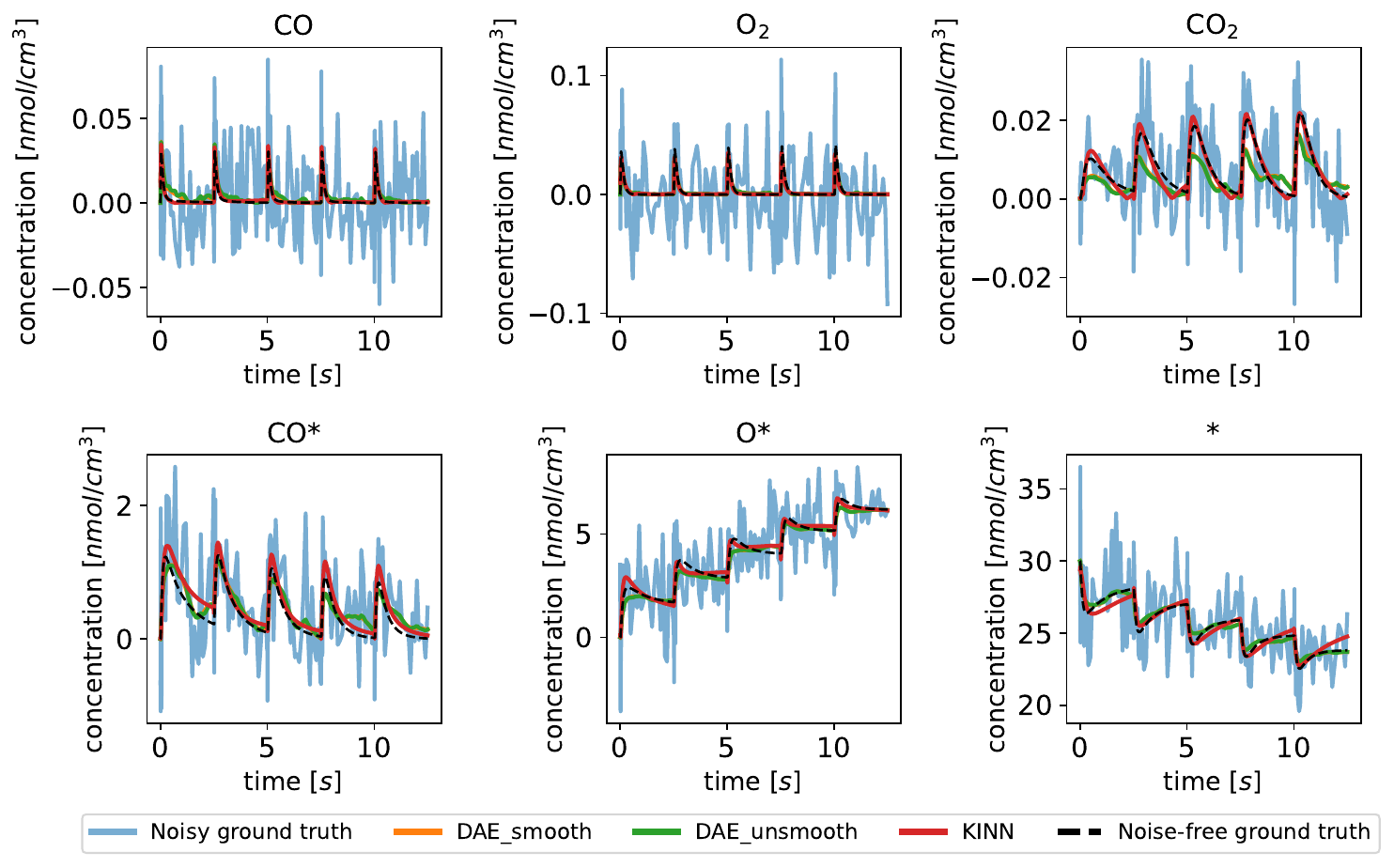}
    \caption{Concentration fitting of DAE and the KINN at 2.0$\sigma$ noise level.}
    \label{fig:dae_20}
\end{figure}

\begin{table}[H]
    \centering
    \begin{tabularx}{\textwidth}{c @{\extracolsep{\fill}} c c c}
    \toprule
        Parameter & KINN & DAE smooth & DAE unsmooth \\
    \midrule
        $k_{1}$   & $10.90$  & $11.60$  & $11.63$  \\
        $k_{-1}$  & $0.129$  & $0.503$  & $0.593$  \\
        $k_{2}$   & $0.348$  & $0.264$  & $0.263$  \\
        $k_{3}$   & $0.434$  & $0.625$  & $0.630$  \\
        $k_{-3}$  & $0.020$  & $0.115$  & $0.117$  \\
        $k_{4}$   & $13.36$  & $10.32$  & $10.19$  \\
    \bottomrule
    \end{tabularx}
    \caption{Kinetic parameters extracted from KINNs and DAE on $0.5\sigma$ noise level.}
    \label{tab:kinn_pyomo}
\end{table}

\begin{table}[H]
    \centering
    \begin{tabularx}{\textwidth}{c @{\extracolsep{\fill}} c c c}
    \toprule
        Parameter & KINN & DAE smooth & DAE unsmooth \\
    \midrule
        $k_{1}$   & $9.587$  & $10.37$  & $10.40$  \\
        $k_{-1}$  & $0.175$  & $0.793$  & $1.009$  \\
        $k_{2}$   & $0.322$  & $0.279$  & $0.277$  \\
        $k_{3}$   & $0.469$  & $19.57$  & $  -  $  \\
        $k_{-3}$  & $0.029$  & $8.069$  & $  -  $  \\
        $k_{4}$   & $10.02$  & $20.92$  & $21.71$  \\
    \bottomrule
    \end{tabularx}
    \caption{Kinetic parameters extracted from KINNs and DAE on $1.0\sigma$ noise level.}
    \label{tab:kinn_pyomo}
\end{table}

\begin{table}[H]
    \centering
    \begin{tabularx}{\textwidth}{c @{\extracolsep{\fill}} c c c}
    \toprule
        Parameter & KINN & DAE smooth & DAE unsmooth \\
    \midrule
        $k_{1}$   & $10.07$  & $9.500$  & $9.647$  \\
        $k_{-1}$  & $0.416$  & $1.354$  & $1.780$  \\
        $k_{2}$   & $0.378$  & $0.312$  & $0.310$  \\
        $k_{3}$   & $0.290$  & $  -  $  & $  -  $  \\
        $k_{-3}$  & $0.021$  & $  -  $  & $  -  $  \\
        $k_{4}$   & $9.761$  & $29.91$  & $32.01$  \\
    \bottomrule
    \end{tabularx}
    \caption{Kinetic parameters extracted from KINNs and DAE on $2.0\sigma$ noise level.}
    \label{tab:kinn_pyomo}
\end{table}

\subsection{Pyomo Setup}

A sample Pyomo script we used to perform the differential algebraic programming comparison is provided. A runnable notebook with data preparation and processing can be found on github.

\begin{lstlisting}[language = Python]
import numpy as np
from pyomo.environ import *
from pyomo.dae import *
import idaes
import json
from scipy.signal import savgol_filter

m = ConcreteModel()
m.t = ContinuousSet(initialize=t)
m.t_meas = Set(initialize=t)

# Define ground truth training data
m.CO_meas  = Param(m.t_meas, initialize=data[0])
m.O2_meas  = Param(m.t_meas, initialize=data[1])
m.CO2_meas = Param(m.t_meas, initialize=data[2])
m.COs_meas = Param(m.t_meas, initialize=data[3])
m.Os_meas  = Param(m.t_meas, initialize=data[4])
m.s_meas   = Param(m.t_meas, initialize=data[5])

# Define net flux for gas species
m.CO_F_meas  = Param(m.t_meas, initialize=F_data[0])
m.O2_F_meas  = Param(m.t_meas, initialize=F_data[1])
m.CO2_F_meas = Param(m.t_meas, initialize=F_data[2])

# Initialize net flux 
m.CO_F  = Var(m.t, initialize=F_data[0])
m.O2_F  = Var(m.t, initialize=F_data[1])
m.CO2_F = Var(m.t, initialize=F_data[2])

# Define concentration terms
m.CO  = Var(m.t, within=NonNegativeReals)
m.O2  = Var(m.t, within=NonNegativeReals)
m.CO2 = Var(m.t, within=NonNegativeReals)
m.COs = Var(m.t, within=NonNegativeReals)
m.Os  = Var(m.t, within=NonNegativeReals)
m.s   = Var(m.t, within=NonNegativeReals)

# Initialize kinetic parameters
m.k1  = Var(initialize = 1e-5, within=PositiveReals)
m.k1_ = Var(initialize = 1e-5, within=PositiveReals)
m.k2  = Var(initialize = 1e-5, within=PositiveReals)
m.k3  = Var(initialize = 1e-5, within=PositiveReals)
m.k3_ = Var(initialize = 1e-5, within=PositiveReals)
m.k4  = Var(initialize = 1e-5, within=PositiveReals)

# dC/dt
m.dCO  = DerivativeVar(m.CO,  wrt=m.t)
m.dO2  = DerivativeVar(m.O2,  wrt=m.t)
m.dCO2 = DerivativeVar(m.CO2, wrt=m.t)
m.dCOs = DerivativeVar(m.COs, wrt=m.t)
m.dOs  = DerivativeVar(m.Os,  wrt=m.t)
m.ds   = DerivativeVar(m.s,   wrt=m.t)

# Define initial conditions
def _init_conditions(m):
    yield m.CO[0.005]  == 1e-5
    yield m.O2[0.005]  == 1e-5
    yield m.CO2[0.005] == 1e-5
    yield m.COs[0.005] == 1e-5
    yield m.Os[0.005]  == 1e-5
    yield m.s[0.005]   == 30.0
m.init_conditions=ConstraintList(rule=_init_conditions)

# Define the kinetic models as function of concentration and flux
def _dCO(m,i):
    if i==0.005:
        return Constraint.Skip
    return m.dCO[i] == - m.k1*m.CO[i]*m.s[i] + m.k1_*m.COs[i] - m.k4*m.CO[i]*m.Os[i] + m.CO_F[i]
m.dCOcon = Constraint(m.t, rule=_dCO)

def _dO2(m,i):
    if i==0.005:
        return Constraint.Skip
    return m.dO2[i] == - m.k2*m.O2[i]*m.s[i]**2 + m.O2_F[i]
m.dO2con = Constraint(m.t, rule=_dO2)

def _dCO2(m,i):
    if i==0.005:
        return Constraint.Skip
    return m.dCO2[i] == m.k3*m.COs[i]*m.Os[i] - m.k3_*m.CO2[i]*m.s[i]**2 + m.k4*m.CO[i]*m.Os[i] + m.CO2_F[i]
m.dCO2con = Constraint(m.t, rule=_dCO2)

def _dCOs(m,i):
    if i==0.005:
        return Constraint.Skip
    return m.dCOs[i] == m.k1*m.CO[i]*m.s[i] - m.k1_*m.COs[i] - m.k3*m.COs[i]*m.Os[i] + m.k3_*m.CO2[i]*m.s[i]**2
m.dCOscon = Constraint(m.t, rule=_dCOs)

def _dOs(m,i):
    if i==0.005:
        return Constraint.Skip
    return m.dOs[i] == (m.k2*m.O2[i]*m.s[i]**2)*2 - m.k3*m.COs[i]*m.Os[i] + m.k3_*m.CO2[i]*m.s[i]**2 - m.k4*m.CO[i]*m.Os[i]
m.dOscon = Constraint(m.t, rule=_dOs)

def _ds(m,i):
    if i==0.005:
        return Constraint.Skip
    return m.ds[i] == -m.k1*m.CO[i]*m.s[i] + m.k1_*m.COs[i] - (m.k2*m.O2[i]*m.s[i]**2)*2 + m.k3*m.COs[i]*m.Os[i]*2 - 2*m.k3_*m.CO2[i]*m.s[i]**2 + m.k4*m.CO[i]*m.Os[i]
m.dscon = Constraint(m.t, rule=_ds)

# Define objective function to minimize the difference between the concentration and flux value and the ground truth
def _obj(m):
    sum1 = sum((m.CO[i]    -    m.CO_meas[i])**2 for i in m.t_meas if i in m.CO_meas.keys()) 
    sum2 = sum((m.O2[i]    -    m.O2_meas[i])**2 for i in m.t_meas if i in m.O2_meas.keys()) * 100      # weighing parameter for scaling
    sum3 = sum((m.CO2[i]   -   m.CO2_meas[i])**2 for i in m.t_meas if i in m.CO2_meas.keys()) * 1000
    sum4 = sum((m.COs[i]   -   m.COs_meas[i])**2 for i in m.t_meas if i in m.COs_meas.keys())
    sum5 = sum((m.Os[i]    -    m.Os_meas[i])**2 for i in m.t_meas if i in m.Os_meas.keys())
    sum6 = sum((m.s[i]     -     m.s_meas[i])**2 for i in m.t_meas if i in m.s_meas.keys())
    sum7 = sum((m.CO_F[i]  -  m.CO_F_meas[i])**2 for i in m.t_meas if i in m.CO_F_meas.keys())
    sum8 = sum((m.O2_F[i]  -  m.O2_F_meas[i])**2 for i in m.t_meas if i in m.O2_F_meas.keys())
    sum9 = sum((m.CO2_F[i] - m.CO2_F_meas[i])**2 for i in m.t_meas if i in m.CO2_F_meas.keys())
    return sum1+sum2+sum3+sum4+sum5+sum6+sum7+sum8+sum9
m.obj = Objective(rule=_obj, sense = minimize)


instance = m.create_instance()

discretizer = TransformationFactory('dae.collocation')
discretizer.apply_to(instance, nfe=30, ncp=1)

solver=SolverFactory('ipopt')
solver.options['max_iter'] = 500000
results = solver.solve(instance,tee=True)
\end{lstlisting}

\section{KINNs Setup}

The KINNs structures and hyperparameter setups are stated in this section. All NN parameters are available on github in $.npz$ format.
\subsection{Single-pulse ideal scenario}

\begin{table}[H]
    \centering
    \begin{tabular}{|l|l|}
    \hline
Activation function                   & swish         \\ \hline
NN structure                          & {[}1, 8, 6{]} \\ \hline
NN parameters initialization scale    & 1e-2          \\ \hline
Model parameters initialization scale & 1e-5          \\ \hline
Step size                             & 1e-3          \\ \hline
Ending $\alpha$                       & 1             \\ \hline
    \end{tabular}
    \label{tab:nn_single}
\end{table}

\subsection{Multi-pulse ideal scenario}

\begin{table}[H]
    \centering
    \begin{tabular}{|l|l|}
    \hline
Activation function                   & swish              \\ \hline
NN structure                          & {[}4, 10, 10, 6{]} \\ \hline
NN parameters initialization scale    & 1e-2               \\ \hline
Model parameters initialization scale & 1e-5               \\ \hline
Step size                             & 1e-3               \\ \hline
Ending $\alpha$                       & 1e-3               \\ \hline
    \end{tabular}
    \label{tab:nn_multi}
\end{table}

\subsection{Multi-pulse practical scenario}

\begin{table}[H]
    \centering
    \begin{tabular}{|l|l|}
    \hline
Activation function                   & swish              \\ \hline
NN structure                          & {[}4, 10, 10, 6{]} \\ \hline
NN parameters initialization scale    & 1e-2               \\ \hline
Model parameters initialization scale & 1e-5               \\ \hline
Step size                             & 1e-3               \\ \hline
Ending $\alpha$                       & 1e-3               \\ \hline
Ending $\beta$                        & 1                  \\ \hline
    \end{tabular}
    \label{tab:nn_multi_Y}
\end{table}